\documentclass[aps,prc,twocolumn,showpacs,floatfix,nofootinbib,preprintnumbers,superscriptaddress,amsmath,amssymb]{revtex4-1}

\usepackage{epsfig}
\usepackage{color}
\usepackage{natbib}
\usepackage{graphicx}
\usepackage{xifthen}

\definecolor{dgreen}{rgb}{0.0,0.6,0.0}

\newcommand{\Si}{$^{34}$Si}
\newcommand{\Ar}{$^{46}$Ar}
\newcommand{\Ca}{$^{48}$Ca}
\newcommand{\Hg}{$^{206}$Hg}
\newcommand{\Pb}{$^{208}$Pb}
\newcommand{\Og}{$^{302}$Og}
\newcommand{\SHE}{$^{472}$164}

\newcommand{\rhomax}[1][]{%
\ifthenelse{\equal{#1}{}}{\rho_\mathrm{max}}{\rho_{#1,\mathrm{max}}}%
}
\newcommand{\rhoav}[1][]{%
\ifthenelse{\equal{#1}{}}{\rho_\mathrm{av}}{\rho_{#1,\mathrm{av}}}%
}
\newcommand{\rhoc}[1][]{%
\ifthenelse{\equal{#1}{}}{\rho_\mathrm{c}}{\rho_{#1,\mathrm{c}}}%
}

\renewcommand{\vec}[1]{\mbox{\boldmath $#1$}}
\newcommand{\vecs}[1]{\mbox{\boldmath \scriptsize$#1$}}

\begin{document}

\title{Central depression in nucleonic densities: Trend analysis
in nuclear density-functional-theory approach}

\author{B. Schuetrumpf}
\affiliation{NSCL/FRIB Laboratory, Michigan State University, East Lansing, Michigan 48824, USA}

\author{W. Nazarewicz}
\affiliation{Department of Physics and Astronomy and FRIB Laboratory, Michigan State University, East Lansing, Michigan 48824, USA}

\author{P.-G.~Reinhard}
\affiliation{Institut f\"ur theoretische Physik, Universit\"at Erlangen, D-91054 Erlangen, Germany}

\date{\today}

\begin{abstract}
\begin{description}
\item[Background]
The central depression of nucleonic density, i.e., a reduction of density in the nuclear interior, has been  attributed to many factors. For instance,  bubble structures in  superheavy nuclei are believed to be due to the electrostatic repulsion. In light nuclei, the mechanism behind the density reduction in the interior  has been discussed in terms  of shell effects associated with  occupations of  $s$-orbits.
\item[Purpose]
The main objective of this work is to reveal mechanisms behind   the formation of central depression in nucleonic densities in light and heavy nuclei. To this end, we introduce several measures of the internal nucleonic density. Through the statistical analysis, we  study the information content of these measures with respect to nuclear matter properties.
\item[Method]
We apply nuclear density functional theory with Skyrme functionals. Using the statistical tools of linear least square  regression, we inspect  correlations between various measures of central depression and model parameters, including nuclear matter properties. We study bivariate correlations with selected quantities as well as multiple correlations with groups of parameters. Detailed correlation analysis is carried out for $^{34}$Si for which a bubble structure has been reported recently, $^{48}$Ca, and $N$=82, 126, and 184 isotonic chains. 
\item[Results]
We show that the central depression in medium-mass nuclei is very sensitive to shell effects, whereas for superheavy systems it is firmly driven by the electrostatic repulsion. An appreciable semi-bubble structure in proton density is  predicted for $^{294}$Og, which is currently the heaviest nucleus known experimentally.
\item[Conclusion]
Our correlation analysis reveals that the central density indicators in nuclei below \Pb{} carry little information on parameters of nuclear matter; they are predominantly driven by shell structure. On the other hand, in the superheavy nuclei there exists a clear relationship between the central nucleonic density and symmetry energy.
\end{description}
\end{abstract}

\maketitle
%===========================================================================
\section{Introduction} The phenomenon of central depression of
nucleonic density, i.e., a reduction of density in the nuclear
interior, has been introduced already in 1946 \cite{Wilson} and
the first quantitative calculations of this effect were performed in
in the early 1970s \cite{DAVIES1972,Campi}. By now, there exists an
appreciable literature devoted to this subject, see, e.g.,
Refs.~\cite{Myers1969,Friedrich1986,Moeller1992,Moretto1996,Royer1996,Pom98,Decharge1999,Bender1999,Berger2001,NazSHE02,Decharge2003,Afa05,Pei2005,Roca-Maza2013,Mehta2015,Todd2004,Grasso2007,Khan2008,Grasso2009,Chu2010,Wang2011,Wang2011a,Liu2012,Yao2012,Nakada2013,Meucci2014,Wang15,Li16,Duguet2017,Mutschler2016}. For
superheavy nuclei, the term ``bubble nucleus" was introduced in the
context of nuclei with vanishing density at the nuclear interior,
  or at least reduced density (semi-bubble). Other exotic
topologies of nucleonic density, such as toroidal configurations
\cite{Siemens,Wong1972,Wong1973,Wong1973a,Wong1977,Warda07,Vinas08,Jac11,Staszczak17}
were also suggested, and calculations of nuclear fragmentation
reactions predicted toroidal and bubble formations
\cite{Borderie1993,Borderie1993a,Bauer1992,Xu1994}.

The appearance of bubble structures in  heavy nuclei has been attributed to
the effect of the electrostatic repulsion by moving protons
towards the nuclear surface. The properties of superheavy bubble
nuclei, including their characteristic  shell structure, have been
studied in, e.g.,
\cite{Myers1969,Friedrich1986,Moeller1992,Moretto1996,Royer1996,Bender1999,Decharge1999,Berger2001,Decharge2003,Afa05,Pei2005,Roca-Maza2013,Mehta2015}. The properties of bubble nuclei can be related to the nuclear equation of
state and the formation of nuclear pasta \cite{Hor08}.

Central depression of nucleonic densities is also expected in light systems such as \Si{} and \Ar{} \cite{DAVIES1972,Todd2004,Grasso2007,Khan2008,Grasso2009,Chu2010,Wang2011,Wang2011a,Liu2012,Yao2012,Nakada2013,Meucci2014,Wu2014,Shukla14,Wang15,Li16,Duguet2017,Mutschler2016}. In contrast to heavy nuclei,  the mechanism behind the density reduction in light systems is related to shell effects. Here, the effect is driven by $s$-orbits, as those are the only states, which contribute to the central density in a non-relativistic picture.
In the case of \Si{} and \Ar{}  it is the vacancy in the proton $1s$
natural orbit that is responsible for the central depression. In heavy nuclei, an excellent candidate  is \Hg{}, where the proton $2s$ natural orbit   is  weakly occupied \cite{Todd2004,Wang15}.

The main objective of this work is to reveal mechanisms behind   the formation of central depression in nucleonic densities in light and heavy nuclei. To this end, we introduce several measures of the internal nucleonic density. Through the statistical analysis, we  study the information content of these measures with respect to nuclear matter properties.

%===========================================================================

\section{Measures of central depression}
A variety of measures of the central depression in nucleonic densities can be found In the literature. A simple and straightforward definition is $(\rhomax-\rhoc)/\rhomax$ \cite{Duguet2017,Yao2012}, where $\rhoc=\rho(\vec{r}=0)$ is the central density and $\rhomax$ is the maximum density. However this quantity is sensitive to oscillations due to shell effects. Additionally it is always positive semi-definite; hence, it cannot quantify the degree of central enhancement, if it is present. To this end, we adopted a slightly different measure:
\begin{equation}
\bar{\rho}_{t,{\rm c}}=({\rhoav[t]-\rhoc[t]})/\rhoav[t],
\end{equation}
where $t=(n,p)$ and $\rhoav[t]=N_t/(4/3 \pi R_d^3)$
is the average density of the nucleus assuming a constant density up
to the diffraction radius $R_\mathrm{d}$ \cite{Fri82a}, also referred to as
box-equivalent radius. We choose $R_\mathrm{d}$ instead of the r.m.s. radius, because this quantity is not affected by the surface thickness.

Another useful indicator of central depression can be obtained from
the charge density form factor, which is a measurable quantity
\cite{Fri82a}. It has been shown that the presence of a central
depression in charge density shifts the zeroes of the form factor
\cite{Friedrich1986,Roca-Maza2013,Chu2010,Liu2012,Meucci2014}.  Within
the modified Helm model \cite{Friedrich1986}, assuming a parabolic
dependence of the density on $r$ around the origin, the central
depression can be parametrized by a dimensionless measure
$\bar{w}_t$. This indicator can be directly obtained from the shift of
the first and second zero of the form factor. The advantage of
$\bar{w}_t$ is that it is fairly robust with respect to shell
fluctuations that predominantly influence the form factor at large
$q$-values \cite{Fri82a}. Positive values of $\bar{w}_t$ correspond to the central
depression while negative values indicate central enhancement.

%===========================================================================

\section{Theoretical framework} 
\subsection{Nuclear DFT}
In order to assess central depression
across the nuclear landscape, we employ nuclear density
functional theory (DFT) \cite{bender2003self} with the globally-optimized Skyrme energy density functionals  SV-min \cite{Kluepfel2009}, SLy6 \cite{Chabanat}, and UNEDF1 \cite{Kortelainen2012b}.  Pairing
is treated at the BCS level.  The pairing space is limited by a soft
cutoff \cite{Bon85a,Kri90a} with the
cutoff parameter  chosen such that it
covers about 1.6 extra oscillator shells above the Fermi energy \cite{Ben00}.
This amounts to a pairing band of about 5 MeV in medium and heavy
nuclei.

\subsection{Correlation analysis}
The results of our DFT calculations are analyzed using the tools
of linear least square regression \cite{Dobaczewski2014}. Our analysis focuses on correlations around the $\chi^2$ minimum of  SV-min. We assume a linear dependence between the model parameters and observables and we checked this assumption {\it a posteriori}. By computing the covariance $\mathrm{cov}(x,y)$ of  quantities $x$ and $y$, as well as their respective variances $\sigma_x$ and $\sigma_y$, we assess
$x$-$y$ correlations  in terms of the bivariate correlation coefficient
\begin{equation}
R_{x,y}=\frac{\mathrm{cov}(x,y)}{\sigma_x\sigma_y}
\end{equation}
or its square $R^2$, which is the coefficient of determination (CoD) \cite{Glantz}. We determine the CoDs as described in Ref.~\cite{Reinhard16}. Note that the
CoD contains information on how well an observable (or model parameter) is determined by another one. However it does not give any information about the associated rate of changes.

Multiple correlation coefficients (MCC) \cite{Allison} of observables
with groups of parameters $\vec{a}$ can determined by computing
\begin{equation}
R^2_{\vecs{a},x}=\vec{c}^T(R_{\vecs{a},\,\vecs{a}})^{-1}\vec{c},
\end{equation}
where $R_{\vecs{a},\vecs{a}}$ is the matrix of CoDs between
the model parameters of group $\vec{a}$ and
$\vec{c}=(R_{a_1,x},R_{a_2,x},...)$ contains the
CoDs between the observables and the single group members. Values of
$R^2$ range from 0 to 1, where 0 implies, that those quantities are
completely uncorrelated, 1 denotes that one quantity determines the
other completely.  An $R^2$ of, say,  $0.30$ means that 30\% of the variance in $x$ is predictable from $\vec{a}$.
For a group containing all model parameters,
an observable is completely determined; hence, $R^2=1$.

\section{Central depression in light and heavy nuclei} 
To avoid the well-known competition between central depression and
shape deformation effects \cite{Afa05,Pei2005,Wu2014}, we will
primarily  consider nuclei that are predicted to be
spherical. Specifically, we study the light- and medium-mass nuclei
\Si{} and \Ca{}, semi-magic isotonic chains $N$=82, 126, and 184, as
well as the the superheavy system \SHE{}.

\begin{figure}[htb]
\includegraphics[width=\linewidth]{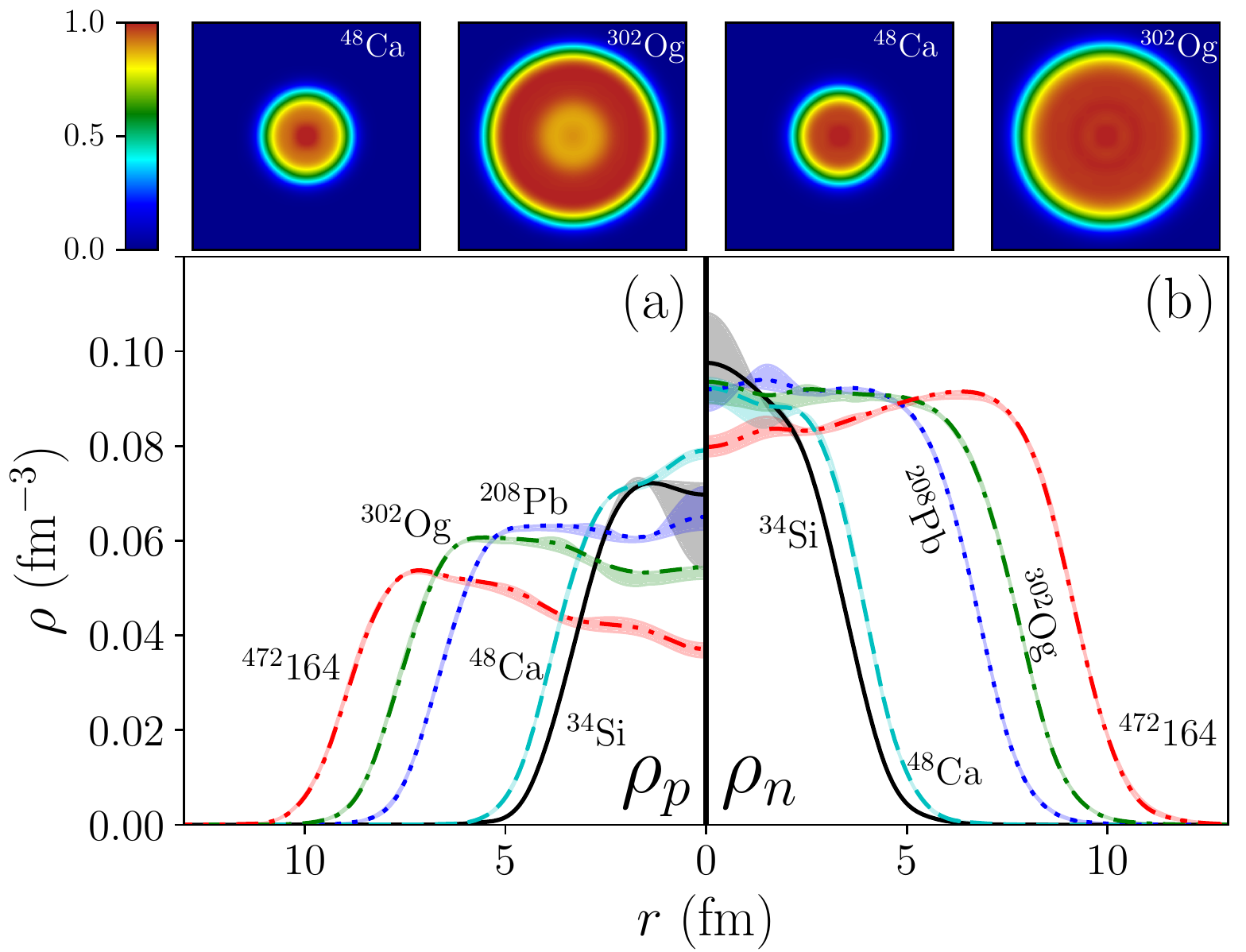}%
\caption{\label{fig:densities} Top: proton (left) and neutron (right)  densities of \Ca{}  and \Og{}, normalized to $\rhomax$,  calculated with SV-min in the ($x,z$) plane at $y=0$. The densities are displayed in a 20\,fm$\times$20\,fm box. Bottom:  proton (left) and neutron (right) densities of \Si{}, \Ca{}, \Pb{}, \Og{}, and \SHE{} obtained with SV-min as functions of $r$. The shaded areas mark the spread of results obtained with  SV-min, SLy6, and UNEDF1.}
\end{figure}

The proton and neutron  densities predicted in SV-min are shown in Fig.~\ref{fig:densities} for several nuclei. It can be seen that superheavy nuclei such as \Og{} and \SHE{} exhibit a pronounced central depression in the proton density distribution. The central depression in \Si{} is predicted to be rather weak by SV-min. The doubly-magic nuclei \Ca{} and \Pb{} show a bump, or enhancement,  in the central proton density. 
The neutron densities displayed in Fig.~\ref{fig:densities}(b) are either flat or exhibit central enhancement. It is only in  \SHE{} that a pronounced central depression in $\rho_n$ is obtained.

The shaded areas indicate the systematic uncertainty stemming from
different choice of a Skyrme functional. The light nucleus \Si{}
exhibits the large uncertainty in the interior. In particular, the
parametrization SLy6 predicts \Si{} to be doubly magic
\cite{Mutschler2016}. The large gap between $0d_{5/2}$ and $1s_{1/2}$
proton shells obtained in this model results in a $1s$-shell vacancy
and large central depression. Other models predict a less
  pronounced  subshell closure at $Z=14$ which results in a
non-vanishing proton pairing, larger $1s_{1/2}$ occupation, and weaker
central depression.  This sensitivity to different models which share
about the same bulk properties suggests that the nature
of central depression in \Si{} is governed by shell effects. This is
consistent with the detailed study of \Si{} in Ref.~\cite{Duguet2017},
which concluded that the ``prediction regarding the (non)existence of
the bubble structure in \Si{} varies significantly with the nuclear
Hamiltonian used." For other nuclei, the systematic uncertainty is
much smaller and SV-min predictions seem to be robust.

\begin{figure}[htb]
\includegraphics[width=\linewidth]{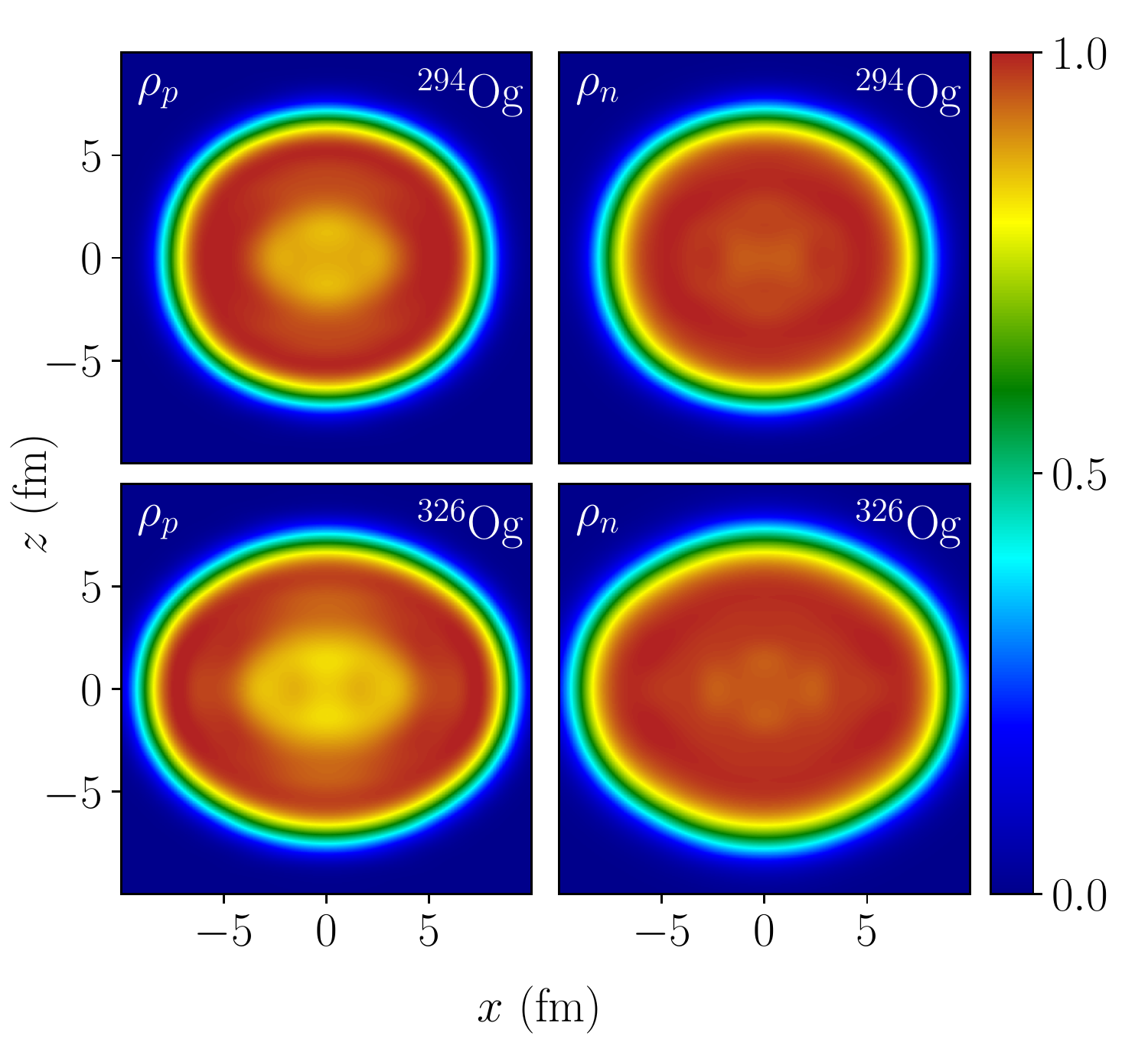}%
\caption{\label{Og-bubbles} Neutron (left) and proton (right)  densities of $^{294}$Og (top)   and  $^{326}$Og (bottom) calculated with SV-min in the ($x,z$) plane at $y=0$. The densities, normalized to $\rhomax$, are displayed in a 20\,fm$\times$20\,fm box.}
\end{figure}

The heaviest nucleus known today is $^{294}$Og \cite{Oga1}. In most DFT calculations \cite{SHENature,Heenen15}, this system is expected to be slightly deformed, with a triaxial shape. To see whether shape deformation can
destroy central depression in $^{294}$Og \cite{Afa05,Pei2005,Wu2014}, in Fig.~\ref{Og-bubbles} we display the proton and neutron  densities in this nucleus, as well as in the heavier 
isotope $^{326}$Og, which is predicted to have an appreciable prolate deformation.
In both cases, the deformed semi-bubble structure in proton density is clearly visible. We can thus conclude that -- according to our calculations --  the region of deformed semi-bubble nuclei has been reached experimentally.

\begin{figure}[htb]
\includegraphics[width=0.8\linewidth]{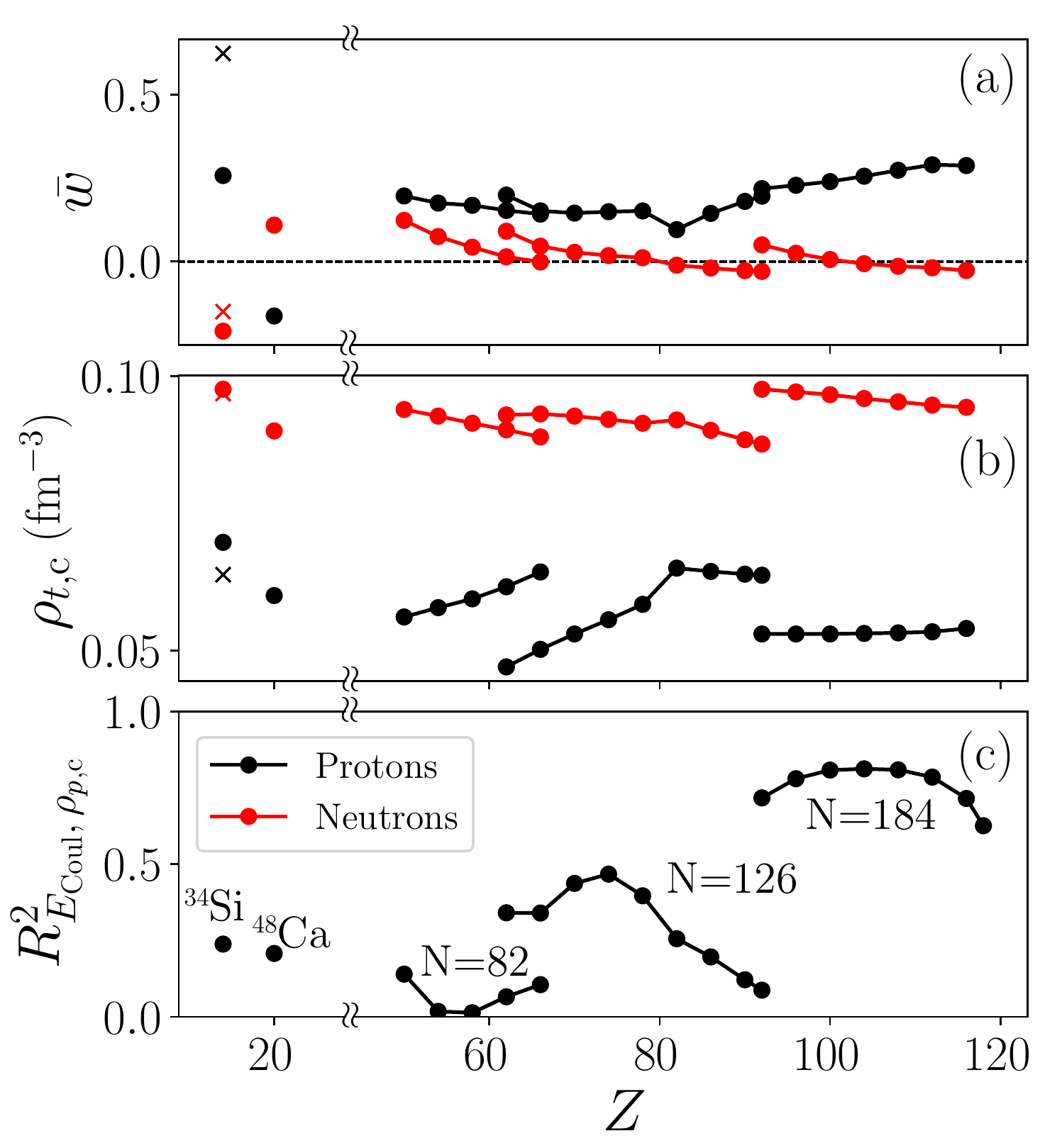}%
\caption{\label{fig:central_depression} Central proton depression  $\bar{w}_{p}$ (a), central  density $\rhoc[t]$ (b), and   CoD between the Coulomb energy $E_\mathrm{Coul}$ and central proton density $\rhoc[p]$ (c) for \Si{}, \Ca{}, and $N=82$,\,126, and 184  isotonic chains predicted by SV-min.
The $\times$ symbol marks the  values of $\bar{w}$ and $\rhoc[t]$ in \Si{} obtained without pairing.}
\end{figure}

Figure~\ref{fig:central_depression}(a) shows the central depression
parameter $\bar{w}_t$ for for \Si{}, \Ca{}, and $N=82$, 126, and 184
isotonic chains predicted by SV-min. As discussed above, the
value of $\bar{w}_p$ in \Si{} predicted in calculations without
pairing increases dramatically.  In heavy and superheavy nuclei, central
proton depression $\bar{w}_p$ is systematically larger than
$\bar{w}_n$. The opposite trend is expected for the central densities
shown Fig.~\ref{fig:central_depression}(b): $\rhoc[p]$ is
systematically reduced as compared to $\rhoc[n]$.

The dip/cusp in \Pb{} can be explained through the full occupation of the $2s$ proton shell, known to be responsible for the central proton depression in $^{206}$Hg. For lighter $N=82$ isotones, the $2s$ shell is partly occupied, e.g., for Pt its occupation is 63\%, and this explains the rise of $\bar{w}_p$ and drop in $\rhoc[p]$. While $\bar{w}_p$ is rather flat for Z$<82$, it smoothly increases with $Z$ along the $N=184$ isotonic chain. This feature is supported by the constant central proton density for the $N=184$ chain seen in Fig.~\ref{fig:central_depression}(b). 

%===========================================================================

\section{Correlation analysis} To understand the origin of trends 
seen in Figs.~\ref{fig:central_depression}(a) and (b), in the following we perform the correlation analysis that  relates the behavior of
key observables related to the central depression  to the parameters of the Skyrme functional.
As relevant observables  we choose the Coulomb energy $E_\mathrm{Coul}$,  central depression parameters $\bar{w}_t$, $\bar{\rho}_{t,{\rm c}}$, $\rhoc[t]$, as well as  the isovector and isoscalar densities at $r=0$: $\rhoc[0]=\rhoc[n]+\rhoc[p]$ and $\rhoc[1]=\rhoc[n]-\rhoc[p]$, respectively.

Figure~\ref{fig:central_depression}(c) displays, in particular, the CoD between $E_\mathrm{Coul}$  and $\rhoc[p]$. It is apparent that for the $N=184$ isotonic chain $\rhoc[p]$ is closely related to $E_\mathrm{Coul}$, whereas for lighter nuclei the correlation between those two parameters is marginal. That results nicely demonstrates that while the central depression  in superheavy nuclei, such as 
the $N=184$ chain, is primarily driven by the electrostatic repulsion, the nature of central depression in lighter systems is different.

While the correlation between  the Coulomb energy and central proton density depression  in superheavy nuclei is apparent, in order to fully understand the origin of central depression one needs to study correlations with the actual Skyrme model parameters. (The Coulomb energy density
functional  primarily depends on the proton density; hence, it cannot be associated with one particular model parameter.)

Some Skyrme functional parameters, characterizing its bulk properties,
can be conveniently expressed through nuclear matter properties (NMP)
in symmetric homogeneous matter. Those are: the equilibrium density
$\rho_{\rm eq}$; energy-per-nucleon at equilibrium $E/A$;
incompressibility $K$; effective mass $m^*/m$ characterizing the
dynamical isoscalar response; symmetry energy $J$; slope of
symmetry energy $L$; and Thomas-Reiche-Kuhn sum-rule enhancement
$\kappa$ characterizing the dynamical isovector response, see
Ref.~\cite{Kluepfel2009,Kortelainen2010} for definitions. In addition,
we consider two parameters characterizing surface properties: surface
energy coefficient $a_\mathrm{surf}$ and surface-symmetry energy
coefficient $a_\mathrm{surf,s}$. Other model parameters, such as those
characterizing spin-orbit and pairing terms yield small correlations
($<20\%$) with the considered observables; hence, they are not
considered in our statistical analysis of CoDs.
%%%%%%%%%%%%
\begin{figure}[htb]
\includegraphics[width=\linewidth]{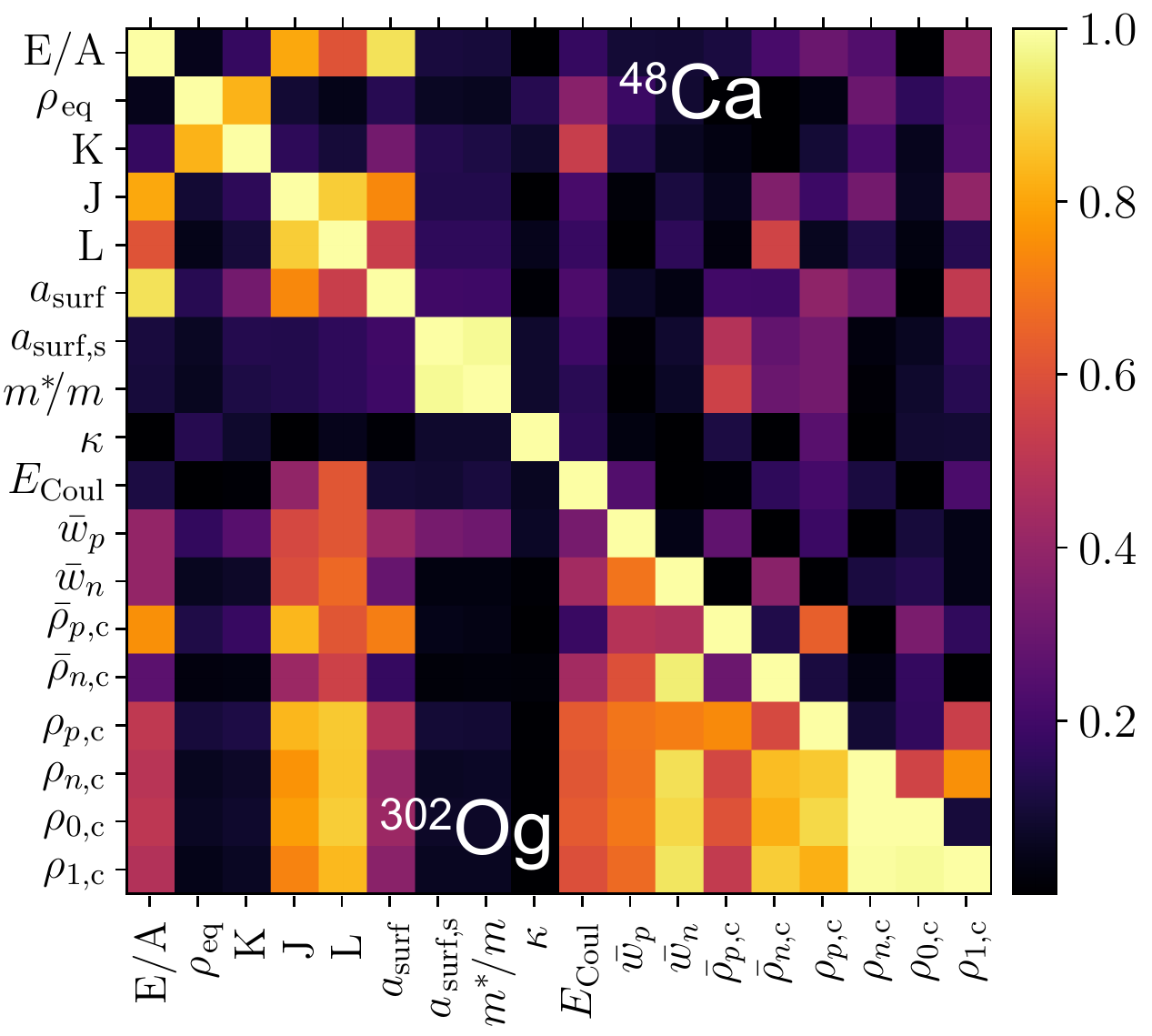}%
\caption{\label{fig:CaOg} Matrices of coefficients of determination for SV-min parameters and selected observables characterizing central densities in \Ca{} (upper triangle) and \Og{} (lower  triangle).}
\end{figure}
 
Figure~\ref{fig:CaOg} shows the matrices of CoDs between the model
parameters and the central density indicators for \Ca{} and \Og{}. The
correlation matrix between the model parameters is nucleus-independent
since it is a property of SV-min parametrization. The
correlations between the different measures of central
  depression are very different for the two nuclei. While the
corresponding CoDs are mostly $<0.5$ for \Ca{} they are appreciable
for \Og{}. This is because the central densities in \Ca{} are
dominated by shell effects, which contribute differently to the
  different measures while global properties dominate in heavy nuclei
  and drive all measures the same way. Furthermore, the correlations
between the model parameters and the central density indicators are
insignificant for \Ca{}, but  show a clear correlation with $E/A$,
  $J$, $L$, and $a_\mathrm{surf}$ for \Og{}.

By studying CoDs for other nuclei we conclude that the central density indicators do not correlate with NMPs for nuclei below \Pb{}. Especially the CoDs for \Pb{} are governed by shell effects, since the exact structure of the $2s$ orbit plays an important role in determining the internal proton density in this nucleus. For nuclei heavier than \Pb{}, the trends seen for \Og{} become more and more pronounced with $Z$. In \SHE{} all central density indicators correlate strongly ($>0.8$).

\begin{figure}[htb]
\includegraphics[width=0.9\linewidth]{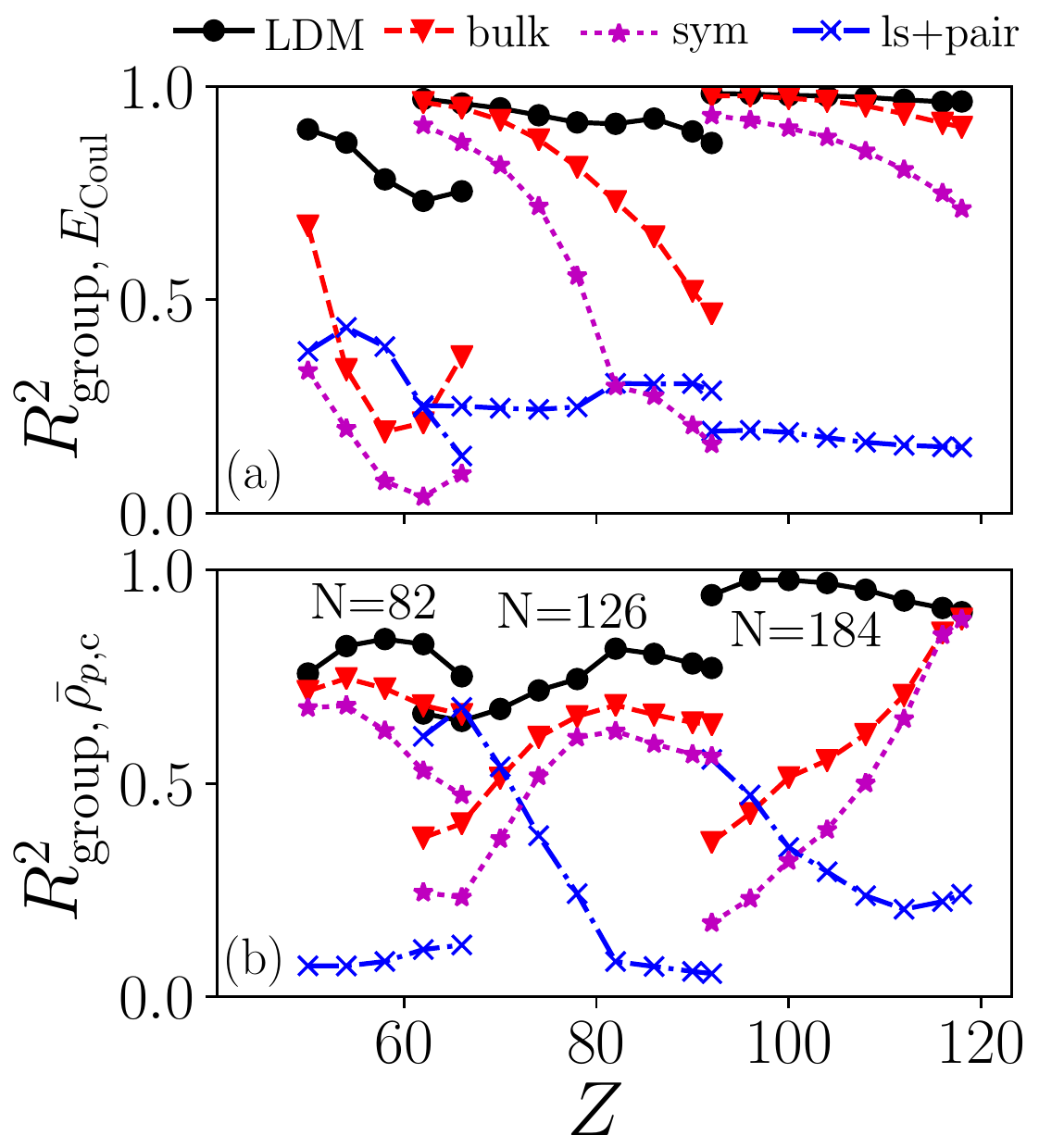}%
\caption{\label{fig:mult} Multiple correlation coefficients with $E_\mathrm{Coul}$ (a) and $\bar{\rho}_{p,{\rm c}}$ (b) in heavy nuclei  with four groups of SV-min parameters:
{\it LDM} ($E/A$, $\rho_\mathrm{eq}$, $K$, $J$, $L$,  $a_\mathrm{surf}$, $a_\mathrm{surf,s}$);  
{\it bulk} ($E/A$, $\rho_\mathrm{eq}$, $K$, $J$, $L$); {\it sym} ($J$, $L$); and
{\it ls+pair} (spin-orbit parameters  $C_0^{\rho \nabla J}$ and  $C_1^{\rho \nabla J}$ and pairing parameters $V_{{\rm pair},n}, V_{{\rm pair},p}$, and $\rho_{\rm pair}$). The surface constants $a_\mathrm{surf}$ and  $a_\mathrm{surf,s}$ are defined as in
Ref.~\cite{Leptodermous}. For other parameters, see Refs.~\cite{Kluepfel2009,Kortelainen2010}.
}
\end{figure}
%%%%%%%%%%%%%%
Correlations with single model parameters can be usefully
complemented by studying MCC.
%Since the model parameters are often correlated (see
%Fig.~\ref{fig:CaOg}), to better understand their impact on observables
%it makes more sense to study 
Figure~\ref{fig:mult} shows MCCs between the four groups of SV-min
parameters and two observables of interest in heavy nuclei: Coulomb
energy and the normalized central proton density $\bar{\rho}_{p,{\rm
    c}}$.  The parameter groups considered here are: liquid drop model
parameters (\textit{LDM}), bulk-properties parameters (\textit{bulk}),
symmetry energy parameters ({\it sym}), and spin-orbit and pairing
parameters (\textit{ls+pair}); see the caption of Fig.~\ref{fig:mult}
for details.

Figure~\ref{fig:mult}(a) illustrates MCCs with the Coulomb energy,
which is closely related to the central depression for heavy nuclei,
see discussion around Fig.~\ref{fig:central_depression}.  The Coulomb
energy is almost entirely determined by {\it LDM}.  The impact of
surface parameters on $E_\mathrm{Coul}$ is large in $N=82$ isotones
while is practically negligible for $N=184$ systems.  Surface effects
tend to increase with $Z$ along the $N=126$ and 184 chains, because
higher charge increases the competition between surface tension and
Coulomb pressure. As expected, the dependence on the symmetry energy
decreases with decreasing isospin/neutron excess.  The {\it ls+pair}
group  of parameters does not impact
  $E_\mathrm{Coul}$ in a meaningful way.

The MCCs with $\bar{\rho}_{p,{\rm c}}$ are shown in
Fig.~\ref{fig:mult}(b). The group correlation with {\it LDM} is
dominant, and increases with $Z$; in superheavy nuclei it becomes
close to 100\%.  The symmetry energy becomes more important for heavy
systems with large isospin where the Coulomb repulsion determines the
central depression. 
(The relevance of the symmetry energy for charge redistribution was  pointed out within the finite-range droplet
model in Refs.~\cite{Myers1969,Moeller1992}.) 
Shell effects impact $\bar{\rho}_{p,{\rm c}}$
weakly for neutron rich nuclei (e.g., above $^{208}$Pb in the
  N=126 chain).  A similar analysis for lighter nuclei \Si{} and
\Ca{} (not shown in Fig.~\ref{fig:mult}) indicates that the relative
contributions from various groups rapidly change from one system to
another.  This, together with large systematic uncertainties for
central densities in \Si{} and \Ca{} shown in
Fig.~\ref{fig:densities}, is indicative of shell-effect dominance on
central density in the low-$Z$ region.

%===========================================================================

\section{Conclusions}
We carried out systematic DFT analysis of the central depression in nucleonic densities in light and heavy nuclei. To study systematic trends of various observables related to internal density we employed statistical tools of linear regression. By inspecting the  coefficients of determination and multiple correlation coefficients we conclude that the central depression of proton density in heavy nuclei is predominantly driven by the LDM parameters. Therein, the origin  of central depression -- resulting in semi-bubble density distributions in superheavy systems --  is  the electrostatic repulsion. 
On the other hand, the central depression appearing in density distributions of lighter nuclei such as \Si{} has its origin in shell effects associated with occupations of $s$-orbits.

The correlation analysis reveals that the central density indicators in nuclei below \Pb{} and especially in \Si{} carry no information on nuclear matter parameters. 
On the other hand, in the superheavy nuclei, which are closer to the leptodermous limit \cite{Leptodermous}, there is a clear relationship between central densities and the symmetry energy.

%======================Acknowledgement=======================================

\begin{acknowledgements}
 This work was supported by the U.S. Department of Energy, Office of Science under Award Numbers DOE-DE-NA0002847 (the Stewardship Science Academic Alliances program), DE-SC0013365 (Michigan State University), and DE-SC0008511 (NUCLEI SciDAC-3 collaboration). An award of computer time was provided by the Institute for Cyber-Enabled Research at Michigan State University.
\end{acknowledgements}
\bibliographystyle{apsrev4-1}
\bibliography{depression}

\begin{thebibliography}{66}%
\makeatletter
\providecommand \@ifxundefined [1]{%
 \@ifx{#1\undefined}
}%
\providecommand \@ifnum [1]{%
 \ifnum #1\expandafter \@firstoftwo
 \else \expandafter \@secondoftwo
 \fi
}%
\providecommand \@ifx [1]{%
 \ifx #1\expandafter \@firstoftwo
 \else \expandafter \@secondoftwo
 \fi
}%
\providecommand \natexlab [1]{#1}%
\providecommand \enquote  [1]{``#1''}%
\providecommand \bibnamefont  [1]{#1}%
\providecommand \bibfnamefont [1]{#1}%
\providecommand \citenamefont [1]{#1}%
\providecommand \href@noop [0]{\@secondoftwo}%
\providecommand \href [0]{\begingroup \@sanitize@url \@href}%
\providecommand \@href[1]{\@@startlink{#1}\@@href}%
\providecommand \@@href[1]{\endgroup#1\@@endlink}%
\providecommand \@sanitize@url [0]{\catcode `\\12\catcode `\$12\catcode
  `\&12\catcode `\#12\catcode `\^12\catcode `\_12\catcode `\%12\relax}%
\providecommand \@@startlink[1]{}%
\providecommand \@@endlink[0]{}%
\providecommand \url  [0]{\begingroup\@sanitize@url \@url }%
\providecommand \@url [1]{\endgroup\@href {#1}{\urlprefix }}%
\providecommand \urlprefix  [0]{URL }%
\providecommand \Eprint [0]{\href }%
\providecommand \doibase [0]{http://dx.doi.org/}%
\providecommand \selectlanguage [0]{\@gobble}%
\providecommand \bibinfo  [0]{\@secondoftwo}%
\providecommand \bibfield  [0]{\@secondoftwo}%
\providecommand \translation [1]{[#1]}%
\providecommand \BibitemOpen [0]{}%
\providecommand \bibitemStop [0]{}%
\providecommand \bibitemNoStop [0]{.\EOS\space}%
\providecommand \EOS [0]{\spacefactor3000\relax}%
\providecommand \BibitemShut  [1]{\csname bibitem#1\endcsname}%
\let\auto@bib@innerbib\@empty
%</preamble>
\bibitem [{\citenamefont {Wilson}(1946)}]{Wilson}%
  \BibitemOpen
  \bibfield  {author} {\bibinfo {author} {\bibfnamefont {H.~A.}\ \bibnamefont
  {Wilson}},\ }\href {\doibase 10.1103/PhysRev.69.538} {\bibfield  {journal}
  {\bibinfo  {journal} {Phys. Rev.}\ }\textbf {\bibinfo {volume} {69}},\
  \bibinfo {pages} {538} (\bibinfo {year} {1946})}\BibitemShut {NoStop}%
\bibitem [{\citenamefont {Davies}\ \emph {et~al.}(1972)\citenamefont {Davies},
  \citenamefont {Wong},\ and\ \citenamefont {Krieger}}]{DAVIES1972}%
  \BibitemOpen
  \bibfield  {author} {\bibinfo {author} {\bibfnamefont {K.}~\bibnamefont
  {Davies}}, \bibinfo {author} {\bibfnamefont {C.}~\bibnamefont {Wong}}, \ and\
  \bibinfo {author} {\bibfnamefont {S.}~\bibnamefont {Krieger}},\ }\href
  {\doibase http://dx.doi.org/10.1016/0370-2693(72)90673-9} {\bibfield
  {journal} {\bibinfo  {journal} {Phys. Lett. B}\ }\textbf {\bibinfo {volume}
  {41}},\ \bibinfo {pages} {455 } (\bibinfo {year} {1972})}\BibitemShut
  {NoStop}%
\bibitem [{\citenamefont {Campi}\ and\ \citenamefont {Sprung}(1973)}]{Campi}%
  \BibitemOpen
  \bibfield  {author} {\bibinfo {author} {\bibfnamefont {X.}~\bibnamefont
  {Campi}}\ and\ \bibinfo {author} {\bibfnamefont {D.}~\bibnamefont {Sprung}},\
  }\href {\doibase http://dx.doi.org/10.1016/0370-2693(73)90121-4} {\bibfield
  {journal} {\bibinfo  {journal} {Phys. Lett. B}\ }\textbf {\bibinfo {volume}
  {46}},\ \bibinfo {pages} {291 } (\bibinfo {year} {1973})}\BibitemShut
  {NoStop}%
\bibitem [{\citenamefont {Myers}\ and\ \citenamefont
  {Swiatecki}(1969)}]{Myers1969}%
  \BibitemOpen
  \bibfield  {author} {\bibinfo {author} {\bibfnamefont {W.}~\bibnamefont
  {Myers}}\ and\ \bibinfo {author} {\bibfnamefont {W.}~\bibnamefont
  {Swiatecki}},\ }\href {\doibase
  http://dx.doi.org/10.1016/0003-4916(69)90202-4} {\bibfield  {journal}
  {\bibinfo  {journal} {Ann. Phys.}\ }\textbf {\bibinfo {volume} {55}},\
  \bibinfo {pages} {395 } (\bibinfo {year} {1969})}\BibitemShut {NoStop}%
\bibitem [{\citenamefont {Friedrich}\ \emph {et~al.}(1986)\citenamefont
  {Friedrich}, \citenamefont {Voegler},\ and\ \citenamefont
  {Reinhard}}]{Friedrich1986}%
  \BibitemOpen
  \bibfield  {author} {\bibinfo {author} {\bibfnamefont {J.}~\bibnamefont
  {Friedrich}}, \bibinfo {author} {\bibfnamefont {N.}~\bibnamefont {Voegler}},
  \ and\ \bibinfo {author} {\bibfnamefont {P.~G.}\ \bibnamefont {Reinhard}},\
  }\href {\doibase 10.1016/0375-9474(86)90053-9} {\bibfield  {journal}
  {\bibinfo  {journal} {Nucl. Phys. A}\ }\textbf {\bibinfo {volume} {459}},\
  \bibinfo {pages} {10} (\bibinfo {year} {1986})}\BibitemShut {NoStop}%
\bibitem [{\citenamefont {M{\"o}ller}\ \emph {et~al.}(1992)\citenamefont
  {M{\"o}ller}, \citenamefont {Nix}, \citenamefont {Myers},\ and\ \citenamefont
  {Swiatecki}}]{Moeller1992}%
  \BibitemOpen
  \bibfield  {author} {\bibinfo {author} {\bibfnamefont {P.}~\bibnamefont
  {M{\"o}ller}}, \bibinfo {author} {\bibfnamefont {J.}~\bibnamefont {Nix}},
  \bibinfo {author} {\bibfnamefont {W.~D.}\ \bibnamefont {Myers}}, \ and\
  \bibinfo {author} {\bibfnamefont {W.~J.}\ \bibnamefont {Swiatecki}},\ }\href
  {\doibase http://dx.doi.org/10.1016/0375-9474(92)90245-F} {\bibfield
  {journal} {\bibinfo  {journal} {Nucl. Phys. A}\ }\textbf {\bibinfo {volume}
  {536}},\ \bibinfo {pages} {61 } (\bibinfo {year} {1992})}\BibitemShut
  {NoStop}%
\bibitem [{\citenamefont {Moretto}\ \emph {et~al.}(1996)\citenamefont
  {Moretto}, \citenamefont {Tso},\ and\ \citenamefont {Wozniak}}]{Moretto1996}%
  \BibitemOpen
  \bibfield  {author} {\bibinfo {author} {\bibfnamefont {L.~G.}\ \bibnamefont
  {Moretto}}, \bibinfo {author} {\bibfnamefont {K.}~\bibnamefont {Tso}}, \ and\
  \bibinfo {author} {\bibfnamefont {G.~J.}\ \bibnamefont {Wozniak}},\
  }\href@noop {} {\bibfield  {journal} {\bibinfo  {journal} {Phys. Rev. Lett.}\
  }\textbf {\bibinfo {volume} {78}},\ \bibinfo {pages} {824} (\bibinfo {year}
  {1996})}\BibitemShut {NoStop}%
\bibitem [{\citenamefont {Royer}\ \emph {et~al.}(1996)\citenamefont {Royer},
  \citenamefont {Haddad},\ and\ \citenamefont {Jouault}}]{Royer1996}%
  \BibitemOpen
  \bibfield  {author} {\bibinfo {author} {\bibfnamefont {G.}~\bibnamefont
  {Royer}}, \bibinfo {author} {\bibfnamefont {F.}~\bibnamefont {Haddad}}, \
  and\ \bibinfo {author} {\bibfnamefont {B.}~\bibnamefont {Jouault}},\ }\href
  {\doibase http://dx.doi.org/10.1016/0375-9474(96)00130-3} {\bibfield
  {journal} {\bibinfo  {journal} {Nucl. Phys. A}\ }\textbf {\bibinfo {volume}
  {605}},\ \bibinfo {pages} {403 } (\bibinfo {year} {1996})}\BibitemShut
  {NoStop}%
\bibitem [{\citenamefont {Dietrich}\ and\ \citenamefont
  {Pomorski}(1998)}]{Pom98}%
  \BibitemOpen
  \bibfield  {author} {\bibinfo {author} {\bibfnamefont {K.}~\bibnamefont
  {Dietrich}}\ and\ \bibinfo {author} {\bibfnamefont {K.}~\bibnamefont
  {Pomorski}},\ }\href {\doibase 10.1103/PhysRevLett.80.37} {\bibfield
  {journal} {\bibinfo  {journal} {Phys. Rev. Lett.}\ }\textbf {\bibinfo
  {volume} {80}},\ \bibinfo {pages} {37} (\bibinfo {year} {1998})}\BibitemShut
  {NoStop}%
\bibitem [{\citenamefont {Decharg{\'e}}\ \emph {et~al.}(1999)\citenamefont
  {Decharg{\'e}}, \citenamefont {Berger}, \citenamefont {Dietrich},\ and\
  \citenamefont {Weiss}}]{Decharge1999}%
  \BibitemOpen
  \bibfield  {author} {\bibinfo {author} {\bibfnamefont {J.}~\bibnamefont
  {Decharg{\'e}}}, \bibinfo {author} {\bibfnamefont {J.-F.}\ \bibnamefont
  {Berger}}, \bibinfo {author} {\bibfnamefont {K.}~\bibnamefont {Dietrich}}, \
  and\ \bibinfo {author} {\bibfnamefont {M.}~\bibnamefont {Weiss}},\ }\href
  {\doibase https://doi.org/10.1016/S0370-2693(99)00225-7} {\bibfield
  {journal} {\bibinfo  {journal} {Phys. Lett. B}\ }\textbf {\bibinfo {volume}
  {451}},\ \bibinfo {pages} {275 } (\bibinfo {year} {1999})}\BibitemShut
  {NoStop}%
\bibitem [{\citenamefont {Bender}(1999)}]{Bender1999}%
  \BibitemOpen
  \bibfield  {author} {\bibinfo {author} {\bibfnamefont {M.}~\bibnamefont
  {Bender}},\ }\href {\doibase 10.1103/PhysRevC.60.034304} {\bibfield
  {journal} {\bibinfo  {journal} {Phys. Rev. C}\ }\textbf {\bibinfo {volume}
  {60}},\ \bibinfo {pages} {34304} (\bibinfo {year} {1999})}\BibitemShut
  {NoStop}%
\bibitem [{\citenamefont {Berger}\ \emph {et~al.}(2001)\citenamefont {Berger},
  \citenamefont {Bitaud}, \citenamefont {Decharg{\'{e}}}, \citenamefont
  {Girod},\ and\ \citenamefont {Dietrich}}]{Berger2001}%
  \BibitemOpen
  \bibfield  {author} {\bibinfo {author} {\bibfnamefont {J.-F.}\ \bibnamefont
  {Berger}}, \bibinfo {author} {\bibfnamefont {L.}~\bibnamefont {Bitaud}},
  \bibinfo {author} {\bibfnamefont {J.}~\bibnamefont {Decharg{\'{e}}}},
  \bibinfo {author} {\bibfnamefont {M.}~\bibnamefont {Girod}}, \ and\ \bibinfo
  {author} {\bibfnamefont {K.}~\bibnamefont {Dietrich}},\ }\href {\doibase
  10.1016/S0375-9474(01)00524-3} {\bibfield  {journal} {\bibinfo  {journal}
  {Nucl. Phys. A}\ }\textbf {\bibinfo {volume} {685}},\ \bibinfo {pages} {1}
  (\bibinfo {year} {2001})}\BibitemShut {NoStop}%
\bibitem [{\citenamefont {Nazarewicz}\ \emph {et~al.}(2002)\citenamefont
  {Nazarewicz}, \citenamefont {Bender}, \citenamefont {\'Cwiok}, \citenamefont
  {Heenen}, \citenamefont {Kruppa}, \citenamefont {Reinhard},\ and\
  \citenamefont {Vertse}}]{NazSHE02}%
  \BibitemOpen
  \bibfield  {author} {\bibinfo {author} {\bibfnamefont {W.}~\bibnamefont
  {Nazarewicz}}, \bibinfo {author} {\bibfnamefont {M.}~\bibnamefont {Bender}},
  \bibinfo {author} {\bibfnamefont {S.}~\bibnamefont {\'Cwiok}}, \bibinfo
  {author} {\bibfnamefont {P.}~\bibnamefont {Heenen}}, \bibinfo {author}
  {\bibfnamefont {A.}~\bibnamefont {Kruppa}}, \bibinfo {author} {\bibfnamefont
  {P.-G.}\ \bibnamefont {Reinhard}}, \ and\ \bibinfo {author} {\bibfnamefont
  {T.}~\bibnamefont {Vertse}},\ }\href {\doibase
  http://dx.doi.org/10.1016/S0375-9474(01)01567-6} {\bibfield  {journal}
  {\bibinfo  {journal} {Nucl. Phys. A}\ }\textbf {\bibinfo {volume} {701}},\
  \bibinfo {pages} {165} (\bibinfo {year} {2002})}\BibitemShut {NoStop}%
\bibitem [{\citenamefont {Decharg{\'{e}}}\ \emph {et~al.}(2003)\citenamefont
  {Decharg{\'{e}}}, \citenamefont {Berger}, \citenamefont {Girod},\ and\
  \citenamefont {Dietrich}}]{Decharge2003}%
  \BibitemOpen
  \bibfield  {author} {\bibinfo {author} {\bibfnamefont {J.}~\bibnamefont
  {Decharg{\'{e}}}}, \bibinfo {author} {\bibfnamefont {J.~F.}\ \bibnamefont
  {Berger}}, \bibinfo {author} {\bibfnamefont {M.}~\bibnamefont {Girod}}, \
  and\ \bibinfo {author} {\bibfnamefont {K.}~\bibnamefont {Dietrich}},\ }\href
  {\doibase 10.1016/S0375-9474(02)01398-2} {\bibfield  {journal} {\bibinfo
  {journal} {Nucl. Phys. A}\ }\textbf {\bibinfo {volume} {716}},\ \bibinfo
  {pages} {55} (\bibinfo {year} {2003})}\BibitemShut {NoStop}%
\bibitem [{\citenamefont {Afanasjev}\ and\ \citenamefont
  {Frauendorf}(2005)}]{Afa05}%
  \BibitemOpen
  \bibfield  {author} {\bibinfo {author} {\bibfnamefont {A.~V.}\ \bibnamefont
  {Afanasjev}}\ and\ \bibinfo {author} {\bibfnamefont {S.}~\bibnamefont
  {Frauendorf}},\ }\href {\doibase 10.1103/PhysRevC.71.024308} {\bibfield
  {journal} {\bibinfo  {journal} {Phys. Rev. C}\ }\textbf {\bibinfo {volume}
  {71}},\ \bibinfo {pages} {024308} (\bibinfo {year} {2005})}\BibitemShut
  {NoStop}%
\bibitem [{\citenamefont {Pei}\ \emph {et~al.}(2005)\citenamefont {Pei},
  \citenamefont {Xu},\ and\ \citenamefont {Stevenson}}]{Pei2005}%
  \BibitemOpen
  \bibfield  {author} {\bibinfo {author} {\bibfnamefont {J.~C.}\ \bibnamefont
  {Pei}}, \bibinfo {author} {\bibfnamefont {F.~R.}\ \bibnamefont {Xu}}, \ and\
  \bibinfo {author} {\bibfnamefont {P.~D.}\ \bibnamefont {Stevenson}},\ }\href
  {\doibase 10.1103/PhysRevC.71.034302} {\bibfield  {journal} {\bibinfo
  {journal} {Phys. Rev. C}\ }\textbf {\bibinfo {volume} {71}},\ \bibinfo
  {pages} {034302} (\bibinfo {year} {2005})}\BibitemShut {NoStop}%
\bibitem [{\citenamefont {Roca-Maza}\ \emph {et~al.}(2013)\citenamefont
  {Roca-Maza}, \citenamefont {Centelles}, \citenamefont {Salvat},\ and\
  \citenamefont {Vi{\~{n}}as}}]{Roca-Maza2013}%
  \BibitemOpen
  \bibfield  {author} {\bibinfo {author} {\bibfnamefont {X.}~\bibnamefont
  {Roca-Maza}}, \bibinfo {author} {\bibfnamefont {M.}~\bibnamefont
  {Centelles}}, \bibinfo {author} {\bibfnamefont {F.}~\bibnamefont {Salvat}}, \
  and\ \bibinfo {author} {\bibfnamefont {X.}~\bibnamefont {Vi{\~{n}}as}},\
  }\href {\doibase 10.1103/PhysRevC.87.014304} {\bibfield  {journal} {\bibinfo
  {journal} {Phy. Rev. C}\ }\textbf {\bibinfo {volume} {87}},\ \bibinfo {pages}
  {014304} (\bibinfo {year} {2013})}\BibitemShut {NoStop}%
\bibitem [{\citenamefont {Mehta}\ \emph {et~al.}(2015)\citenamefont {Mehta},
  \citenamefont {Kaur}, \citenamefont {Kumar},\ and\ \citenamefont
  {Patra}}]{Mehta2015}%
  \BibitemOpen
  \bibfield  {author} {\bibinfo {author} {\bibfnamefont {M.~S.}\ \bibnamefont
  {Mehta}}, \bibinfo {author} {\bibfnamefont {H.}~\bibnamefont {Kaur}},
  \bibinfo {author} {\bibfnamefont {B.}~\bibnamefont {Kumar}}, \ and\ \bibinfo
  {author} {\bibfnamefont {S.~K.}\ \bibnamefont {Patra}},\ }\href {\doibase
  10.1103/PhysRevC.92.054305} {\bibfield  {journal} {\bibinfo  {journal} {Phys.
  Rev. C}\ }\textbf {\bibinfo {volume} {92}},\ \bibinfo {pages} {054305}
  (\bibinfo {year} {2015})}\BibitemShut {NoStop}%
\bibitem [{\citenamefont {Todd-Rutel}\ \emph {et~al.}(2004)\citenamefont
  {Todd-Rutel}, \citenamefont {Piekarewicz},\ and\ \citenamefont
  {Cottle}}]{Todd2004}%
  \BibitemOpen
  \bibfield  {author} {\bibinfo {author} {\bibfnamefont {B.~G.}\ \bibnamefont
  {Todd-Rutel}}, \bibinfo {author} {\bibfnamefont {J.}~\bibnamefont
  {Piekarewicz}}, \ and\ \bibinfo {author} {\bibfnamefont {P.~D.}\ \bibnamefont
  {Cottle}},\ }\href {\doibase 10.1103/PhysRevC.69.021301} {\bibfield
  {journal} {\bibinfo  {journal} {Phys. Rev. C}\ }\textbf {\bibinfo {volume}
  {69}},\ \bibinfo {pages} {021301} (\bibinfo {year} {2004})}\BibitemShut
  {NoStop}%
\bibitem [{\citenamefont {Grasso}\ \emph {et~al.}(2007)\citenamefont {Grasso},
  \citenamefont {Ma}, \citenamefont {Khan}, \citenamefont {Margueron},\ and\
  \citenamefont {Giai}}]{Grasso2007}%
  \BibitemOpen
  \bibfield  {author} {\bibinfo {author} {\bibfnamefont {M.}~\bibnamefont
  {Grasso}}, \bibinfo {author} {\bibfnamefont {Z.~Y.}\ \bibnamefont {Ma}},
  \bibinfo {author} {\bibfnamefont {E.}~\bibnamefont {Khan}}, \bibinfo {author}
  {\bibfnamefont {J.}~\bibnamefont {Margueron}}, \ and\ \bibinfo {author}
  {\bibfnamefont {N.~V.}\ \bibnamefont {Giai}},\ }\href {\doibase
  10.1103/PhysRevC.76.044319} {\bibfield  {journal} {\bibinfo  {journal} {Phys.
  Rev. C}\ }\textbf {\bibinfo {volume} {76}},\ \bibinfo {pages} {044319}
  (\bibinfo {year} {2007})}\BibitemShut {NoStop}%
\bibitem [{\citenamefont {Khan}\ \emph {et~al.}(2008)\citenamefont {Khan},
  \citenamefont {Grasso}, \citenamefont {Margueron},\ and\ \citenamefont
  {Giai}}]{Khan2008}%
  \BibitemOpen
  \bibfield  {author} {\bibinfo {author} {\bibfnamefont {E.}~\bibnamefont
  {Khan}}, \bibinfo {author} {\bibfnamefont {M.}~\bibnamefont {Grasso}},
  \bibinfo {author} {\bibfnamefont {J.}~\bibnamefont {Margueron}}, \ and\
  \bibinfo {author} {\bibfnamefont {N.~V.}\ \bibnamefont {Giai}},\ }\href
  {\doibase http://dx.doi.org/10.1016/j.nuclphysa.2007.11.012} {\bibfield
  {journal} {\bibinfo  {journal} {Nucl. Phys. A}\ }\textbf {\bibinfo {volume}
  {800}},\ \bibinfo {pages} {37 } (\bibinfo {year} {2008})}\BibitemShut
  {NoStop}%
\bibitem [{\citenamefont {Grasso}\ \emph {et~al.}(2009)\citenamefont {Grasso},
  \citenamefont {Gaudefroy}, \citenamefont {Khan}, \citenamefont
  {Nik{\v{s}}i{\'{c}}}, \citenamefont {Piekarewicz}, \citenamefont {Sorlin},
  \citenamefont {Giai},\ and\ \citenamefont {Vretenar}}]{Grasso2009}%
  \BibitemOpen
  \bibfield  {author} {\bibinfo {author} {\bibfnamefont {M.}~\bibnamefont
  {Grasso}}, \bibinfo {author} {\bibfnamefont {L.}~\bibnamefont {Gaudefroy}},
  \bibinfo {author} {\bibfnamefont {E.}~\bibnamefont {Khan}}, \bibinfo {author}
  {\bibfnamefont {T.}~\bibnamefont {Nik{\v{s}}i{\'{c}}}}, \bibinfo {author}
  {\bibfnamefont {J.}~\bibnamefont {Piekarewicz}}, \bibinfo {author}
  {\bibfnamefont {O.}~\bibnamefont {Sorlin}}, \bibinfo {author} {\bibfnamefont
  {N.~V.}\ \bibnamefont {Giai}}, \ and\ \bibinfo {author} {\bibfnamefont
  {D.}~\bibnamefont {Vretenar}},\ }\href {\doibase 10.1103/PhysRevC.79.034318}
  {\bibfield  {journal} {\bibinfo  {journal} {Phys. Rev. C}\ }\textbf {\bibinfo
  {volume} {79}},\ \bibinfo {pages} {034318} (\bibinfo {year}
  {2009})}\BibitemShut {NoStop}%
\bibitem [{\citenamefont {Chu}\ \emph {et~al.}(2010)\citenamefont {Chu},
  \citenamefont {Ren}, \citenamefont {Wang},\ and\ \citenamefont
  {Dong}}]{Chu2010}%
  \BibitemOpen
  \bibfield  {author} {\bibinfo {author} {\bibfnamefont {Y.}~\bibnamefont
  {Chu}}, \bibinfo {author} {\bibfnamefont {Z.}~\bibnamefont {Ren}}, \bibinfo
  {author} {\bibfnamefont {Z.}~\bibnamefont {Wang}}, \ and\ \bibinfo {author}
  {\bibfnamefont {T.}~\bibnamefont {Dong}},\ }\href {\doibase
  10.1103/PhysRevC.82.024320} {\bibfield  {journal} {\bibinfo  {journal} {Phys.
  Rev. C}\ }\textbf {\bibinfo {volume} {82}},\ \bibinfo {pages} {024320}
  (\bibinfo {year} {2010})}\BibitemShut {NoStop}%
\bibitem [{\citenamefont {Wang}\ \emph {et~al.}(2011)\citenamefont {Wang},
  \citenamefont {Gu}, \citenamefont {Zhang},\ and\ \citenamefont
  {Dong}}]{Wang2011}%
  \BibitemOpen
  \bibfield  {author} {\bibinfo {author} {\bibfnamefont {Y.~Z.}\ \bibnamefont
  {Wang}}, \bibinfo {author} {\bibfnamefont {J.~Z.}\ \bibnamefont {Gu}},
  \bibinfo {author} {\bibfnamefont {X.~Z.}\ \bibnamefont {Zhang}}, \ and\
  \bibinfo {author} {\bibfnamefont {J.~M.}\ \bibnamefont {Dong}},\ }\href
  {\doibase 10.1103/PhysRevC.84.044333} {\bibfield  {journal} {\bibinfo
  {journal} {Phys. Rev. C}\ }\textbf {\bibinfo {volume} {84}},\ \bibinfo
  {pages} {044333} (\bibinfo {year} {2011})}\BibitemShut {NoStop}%
\bibitem [{\citenamefont {Yan-Zhao}\ \emph {et~al.}(2011)\citenamefont
  {Yan-Zhao}, \citenamefont {Jian-Zhong}, \citenamefont {Xi-Zhen},\ and\
  \citenamefont {Jian-Min}}]{Wang2011a}%
  \BibitemOpen
  \bibfield  {author} {\bibinfo {author} {\bibfnamefont {W.}~\bibnamefont
  {Yan-Zhao}}, \bibinfo {author} {\bibfnamefont {G.}~\bibnamefont
  {Jian-Zhong}}, \bibinfo {author} {\bibfnamefont {Z.}~\bibnamefont {Xi-Zhen}},
  \ and\ \bibinfo {author} {\bibfnamefont {D.}~\bibnamefont {Jian-Min}},\
  }\href {http://stacks.iop.org/0256-307X/28/i=10/a=102101} {\bibfield
  {journal} {\bibinfo  {journal} {Chin. Phys. Lett.}\ }\textbf {\bibinfo
  {volume} {28}},\ \bibinfo {pages} {102101} (\bibinfo {year}
  {2011})}\BibitemShut {NoStop}%
\bibitem [{\citenamefont {Liu}\ \emph {et~al.}(2012)\citenamefont {Liu},
  \citenamefont {Chu}, \citenamefont {Ren},\ and\ \citenamefont
  {Wang}}]{Liu2012}%
  \BibitemOpen
  \bibfield  {author} {\bibinfo {author} {\bibfnamefont {J.}~\bibnamefont
  {Liu}}, \bibinfo {author} {\bibfnamefont {Y.-Y.}\ \bibnamefont {Chu}},
  \bibinfo {author} {\bibfnamefont {Z.-Z.}\ \bibnamefont {Ren}}, \ and\
  \bibinfo {author} {\bibfnamefont {Z.-J.}\ \bibnamefont {Wang}},\ }\href
  {\doibase 10.1088/1674-1137/36/1/008} {\bibfield  {journal} {\bibinfo
  {journal} {Chin. Phys. C}\ }\textbf {\bibinfo {volume} {36}},\ \bibinfo
  {pages} {48} (\bibinfo {year} {2012})}\BibitemShut {NoStop}%
\bibitem [{\citenamefont {Yao}\ \emph {et~al.}(2012)\citenamefont {Yao},
  \citenamefont {Baroni}, \citenamefont {Bender},\ and\ \citenamefont
  {Heenen}}]{Yao2012}%
  \BibitemOpen
  \bibfield  {author} {\bibinfo {author} {\bibfnamefont {J.-M.}\ \bibnamefont
  {Yao}}, \bibinfo {author} {\bibfnamefont {S.}~\bibnamefont {Baroni}},
  \bibinfo {author} {\bibfnamefont {M.}~\bibnamefont {Bender}}, \ and\ \bibinfo
  {author} {\bibfnamefont {P.-H.}\ \bibnamefont {Heenen}},\ }\href {\doibase
  10.1103/PhysRevC.86.014310} {\bibfield  {journal} {\bibinfo  {journal} {Phys.
  Rev. C}\ }\textbf {\bibinfo {volume} {86}},\ \bibinfo {pages} {014310}
  (\bibinfo {year} {2012})}\BibitemShut {NoStop}%
\bibitem [{\citenamefont {Nakada}\ \emph {et~al.}(2013)\citenamefont {Nakada},
  \citenamefont {Sugiura},\ and\ \citenamefont {Margueron}}]{Nakada2013}%
  \BibitemOpen
  \bibfield  {author} {\bibinfo {author} {\bibfnamefont {H.}~\bibnamefont
  {Nakada}}, \bibinfo {author} {\bibfnamefont {K.}~\bibnamefont {Sugiura}}, \
  and\ \bibinfo {author} {\bibfnamefont {J.}~\bibnamefont {Margueron}},\ }\href
  {\doibase 10.1103/PhysRevC.87.067305} {\bibfield  {journal} {\bibinfo
  {journal} {Phys. Rev. C}\ }\textbf {\bibinfo {volume} {87}},\ \bibinfo
  {pages} {067305} (\bibinfo {year} {2013})}\BibitemShut {NoStop}%
\bibitem [{\citenamefont {Meucci}\ \emph {et~al.}(2014)\citenamefont {Meucci},
  \citenamefont {Vorabbi}, \citenamefont {Giusti}, \citenamefont {Pacati},\
  and\ \citenamefont {Finelli}}]{Meucci2014}%
  \BibitemOpen
  \bibfield  {author} {\bibinfo {author} {\bibfnamefont {A.}~\bibnamefont
  {Meucci}}, \bibinfo {author} {\bibfnamefont {M.}~\bibnamefont {Vorabbi}},
  \bibinfo {author} {\bibfnamefont {C.}~\bibnamefont {Giusti}}, \bibinfo
  {author} {\bibfnamefont {F.~D.}\ \bibnamefont {Pacati}}, \ and\ \bibinfo
  {author} {\bibfnamefont {P.}~\bibnamefont {Finelli}},\ }\href {\doibase
  10.1103/PhysRevC.89.034604} {\bibfield  {journal} {\bibinfo  {journal} {Phys
  Rev. C}\ }\textbf {\bibinfo {volume} {89}},\ \bibinfo {pages} {034604}
  (\bibinfo {year} {2014})}\BibitemShut {NoStop}%
\bibitem [{\citenamefont {Wang}\ \emph {et~al.}(2015)\citenamefont {Wang},
  \citenamefont {Hou}, \citenamefont {Zhang}, \citenamefont {Tian},\ and\
  \citenamefont {Gu}}]{Wang15}%
  \BibitemOpen
  \bibfield  {author} {\bibinfo {author} {\bibfnamefont {Y.~Z.}\ \bibnamefont
  {Wang}}, \bibinfo {author} {\bibfnamefont {Z.~Y.}\ \bibnamefont {Hou}},
  \bibinfo {author} {\bibfnamefont {Q.~L.}\ \bibnamefont {Zhang}}, \bibinfo
  {author} {\bibfnamefont {R.~L.}\ \bibnamefont {Tian}}, \ and\ \bibinfo
  {author} {\bibfnamefont {J.~Z.}\ \bibnamefont {Gu}},\ }\href {\doibase
  10.1103/PhysRevC.91.017302} {\bibfield  {journal} {\bibinfo  {journal} {Phys.
  Rev. C}\ }\textbf {\bibinfo {volume} {91}},\ \bibinfo {pages} {017302}
  (\bibinfo {year} {2015})}\BibitemShut {NoStop}%
\bibitem [{\citenamefont {Li}\ \emph {et~al.}(2016)\citenamefont {Li},
  \citenamefont {Long}, \citenamefont {Song},\ and\ \citenamefont
  {Zhao}}]{Li16}%
  \BibitemOpen
  \bibfield  {author} {\bibinfo {author} {\bibfnamefont {J.~J.}\ \bibnamefont
  {Li}}, \bibinfo {author} {\bibfnamefont {W.~H.}\ \bibnamefont {Long}},
  \bibinfo {author} {\bibfnamefont {J.~L.}\ \bibnamefont {Song}}, \ and\
  \bibinfo {author} {\bibfnamefont {Q.}~\bibnamefont {Zhao}},\ }\href {\doibase
  10.1103/PhysRevC.93.054312} {\bibfield  {journal} {\bibinfo  {journal} {Phys.
  Rev. C}\ }\textbf {\bibinfo {volume} {93}},\ \bibinfo {pages} {054312}
  (\bibinfo {year} {2016})}\BibitemShut {NoStop}%
\bibitem [{\citenamefont {Duguet}\ \emph {et~al.}(2017)\citenamefont {Duguet},
  \citenamefont {Som{\`{a}}}, \citenamefont {Lecluse}, \citenamefont
  {Barbieri},\ and\ \citenamefont {Navr{\'{a}}til}}]{Duguet2017}%
  \BibitemOpen
  \bibfield  {author} {\bibinfo {author} {\bibfnamefont {T.}~\bibnamefont
  {Duguet}}, \bibinfo {author} {\bibfnamefont {V.}~\bibnamefont {Som{\`{a}}}},
  \bibinfo {author} {\bibfnamefont {S.}~\bibnamefont {Lecluse}}, \bibinfo
  {author} {\bibfnamefont {C.}~\bibnamefont {Barbieri}}, \ and\ \bibinfo
  {author} {\bibfnamefont {P.}~\bibnamefont {Navr{\'{a}}til}},\ }\href
  {\doibase 10.1103/PhysRevC.95.034319} {\bibfield  {journal} {\bibinfo
  {journal} {Phys. Rev. C}\ }\textbf {\bibinfo {volume} {95}},\ \bibinfo
  {pages} {034319} (\bibinfo {year} {2017})}\BibitemShut {NoStop}%
\bibitem [{\citenamefont {Mutschler}\ \emph {et~al.}(2016)\citenamefont
  {Mutschler}, \citenamefont {Lemasson}, \citenamefont {Sorlin}, \citenamefont
  {Bazin}, \citenamefont {Borcea}, \citenamefont {Borcea}, \citenamefont
  {Dombr{\'{a}}di}, \citenamefont {Ebran}, \citenamefont {Gade}, \citenamefont
  {Iwasaki}, \citenamefont {Khan}, \citenamefont {Lepailleur}, \citenamefont
  {Recchia}, \citenamefont {Roger}, \citenamefont {Rotaru}, \citenamefont
  {Sohler}, \citenamefont {Stanoiu}, \citenamefont {Stroberg}, \citenamefont
  {Tostevin}, \citenamefont {Vandebrouck}, \citenamefont {Weisshaar},\ and\
  \citenamefont {Wimmer}}]{Mutschler2016}%
  \BibitemOpen
  \bibfield  {author} {\bibinfo {author} {\bibfnamefont {A.}~\bibnamefont
  {Mutschler}}, \bibinfo {author} {\bibfnamefont {A.}~\bibnamefont {Lemasson}},
  \bibinfo {author} {\bibfnamefont {O.}~\bibnamefont {Sorlin}}, \bibinfo
  {author} {\bibfnamefont {D.}~\bibnamefont {Bazin}}, \bibinfo {author}
  {\bibfnamefont {C.}~\bibnamefont {Borcea}}, \bibinfo {author} {\bibfnamefont
  {R.}~\bibnamefont {Borcea}}, \bibinfo {author} {\bibfnamefont
  {Z.}~\bibnamefont {Dombr{\'{a}}di}}, \bibinfo {author} {\bibfnamefont
  {J.-P.}\ \bibnamefont {Ebran}}, \bibinfo {author} {\bibfnamefont
  {A.}~\bibnamefont {Gade}}, \bibinfo {author} {\bibfnamefont {H.}~\bibnamefont
  {Iwasaki}}, \bibinfo {author} {\bibfnamefont {E.}~\bibnamefont {Khan}},
  \bibinfo {author} {\bibfnamefont {A.}~\bibnamefont {Lepailleur}}, \bibinfo
  {author} {\bibfnamefont {F.}~\bibnamefont {Recchia}}, \bibinfo {author}
  {\bibfnamefont {T.}~\bibnamefont {Roger}}, \bibinfo {author} {\bibfnamefont
  {F.}~\bibnamefont {Rotaru}}, \bibinfo {author} {\bibfnamefont
  {D.}~\bibnamefont {Sohler}}, \bibinfo {author} {\bibfnamefont
  {M.}~\bibnamefont {Stanoiu}}, \bibinfo {author} {\bibfnamefont {S.~R.}\
  \bibnamefont {Stroberg}}, \bibinfo {author} {\bibfnamefont {J.~A.}\
  \bibnamefont {Tostevin}}, \bibinfo {author} {\bibfnamefont {M.}~\bibnamefont
  {Vandebrouck}}, \bibinfo {author} {\bibfnamefont {D.}~\bibnamefont
  {Weisshaar}}, \ and\ \bibinfo {author} {\bibfnamefont {K.}~\bibnamefont
  {Wimmer}},\ }\href {\doibase 10.1038/nphys3916} {\bibfield  {journal}
  {\bibinfo  {journal} {Nature Phys.}\ }\textbf {\bibinfo {volume} {13}},\
  \bibinfo {pages} {152} (\bibinfo {year} {2016})}\BibitemShut {NoStop}%
\bibitem [{\citenamefont {Siemens}\ and\ \citenamefont
  {Bethe}(1967)}]{Siemens}%
  \BibitemOpen
  \bibfield  {author} {\bibinfo {author} {\bibfnamefont {P.~J.}\ \bibnamefont
  {Siemens}}\ and\ \bibinfo {author} {\bibfnamefont {H.~A.}\ \bibnamefont
  {Bethe}},\ }\href {\doibase 10.1103/PhysRevLett.18.704} {\bibfield  {journal}
  {\bibinfo  {journal} {Phys. Rev. Lett.}\ }\textbf {\bibinfo {volume} {18}},\
  \bibinfo {pages} {704} (\bibinfo {year} {1967})}\BibitemShut {NoStop}%
\bibitem [{\citenamefont {Wong}(1972)}]{Wong1972}%
  \BibitemOpen
  \bibfield  {author} {\bibinfo {author} {\bibfnamefont {C.}~\bibnamefont
  {Wong}},\ }\href {\doibase http://dx.doi.org/10.1016/0370-2693(72)90671-5}
  {\bibfield  {journal} {\bibinfo  {journal} {Phys. Lett. B}\ }\textbf
  {\bibinfo {volume} {41}},\ \bibinfo {pages} {446 } (\bibinfo {year}
  {1972})}\BibitemShut {NoStop}%
\bibitem [{\citenamefont {Wong}(1985)}]{Wong1973}%
  \BibitemOpen
  \bibfield  {author} {\bibinfo {author} {\bibfnamefont {C.-Y.}\ \bibnamefont
  {Wong}},\ }\href {\doibase 10.1103/PhysRevLett.55.1973} {\bibfield  {journal}
  {\bibinfo  {journal} {Phys. Rev. Lett.}\ }\textbf {\bibinfo {volume} {55}},\
  \bibinfo {pages} {1973} (\bibinfo {year} {1985})}\BibitemShut {NoStop}%
\bibitem [{\citenamefont {Wong}(1973)}]{Wong1973a}%
  \BibitemOpen
  \bibfield  {author} {\bibinfo {author} {\bibfnamefont {C.}~\bibnamefont
  {Wong}},\ }\href {\doibase http://dx.doi.org/10.1016/0003-4916(73)90420-X}
  {\bibfield  {journal} {\bibinfo  {journal} {Ann. Phys.}\ }\textbf {\bibinfo
  {volume} {77}},\ \bibinfo {pages} {279 } (\bibinfo {year}
  {1973})}\BibitemShut {NoStop}%
\bibitem [{\citenamefont {Wong}\ \emph {et~al.}(1977)\citenamefont {Wong},
  \citenamefont {Maruhn},\ and\ \citenamefont {Welton}}]{Wong1977}%
  \BibitemOpen
  \bibfield  {author} {\bibinfo {author} {\bibfnamefont {C.}~\bibnamefont
  {Wong}}, \bibinfo {author} {\bibfnamefont {J.}~\bibnamefont {Maruhn}}, \ and\
  \bibinfo {author} {\bibfnamefont {T.}~\bibnamefont {Welton}},\ }\href
  {\doibase http://dx.doi.org/10.1016/0370-2693(77)90602-5} {\bibfield
  {journal} {\bibinfo  {journal} {Phys. Lett. B}\ }\textbf {\bibinfo {volume}
  {66}},\ \bibinfo {pages} {19 } (\bibinfo {year} {1977})}\BibitemShut
  {NoStop}%
\bibitem [{\citenamefont {Warda}(2008)}]{Warda07}%
  \BibitemOpen
  \bibfield  {author} {\bibinfo {author} {\bibfnamefont {M.}~\bibnamefont
  {Warda}},\ }\href
  {http://www.worldscientific.com/doi/abs/10.1142/S0218301307005880} {\bibfield
   {journal} {\bibinfo  {journal} {Int. J. Mod. Phys. E}\ }\textbf {\bibinfo
  {volume} {16}},\ \bibinfo {pages} {452} (\bibinfo {year} {2008})}\BibitemShut
  {NoStop}%
\bibitem [{\citenamefont {Vi\~{n}as}\ \emph {et~al.}(2008)\citenamefont
  {Vi\~{n}as}, \citenamefont {Centelles},\ and\ \citenamefont
  {Warda}}]{Vinas08}%
  \BibitemOpen
  \bibfield  {author} {\bibinfo {author} {\bibfnamefont {X.}~\bibnamefont
  {Vi\~{n}as}}, \bibinfo {author} {\bibfnamefont {M.}~\bibnamefont
  {Centelles}}, \ and\ \bibinfo {author} {\bibfnamefont {M.}~\bibnamefont
  {Warda}},\ }\href {\doibase 10.1142/S0218301308009677} {\bibfield  {journal}
  {\bibinfo  {journal} {Int. J. Mod. Phys. E}\ }\textbf {\bibinfo {volume}
  {17}},\ \bibinfo {pages} {177} (\bibinfo {year} {2008})}\BibitemShut
  {NoStop}%
\bibitem [{\citenamefont {Jachimowicz}\ \emph {et~al.}(2011)\citenamefont
  {Jachimowicz}, \citenamefont {Kowal},\ and\ \citenamefont {Skalski}}]{Jac11}%
  \BibitemOpen
  \bibfield  {author} {\bibinfo {author} {\bibfnamefont {P.}~\bibnamefont
  {Jachimowicz}}, \bibinfo {author} {\bibfnamefont {M.}~\bibnamefont {Kowal}},
  \ and\ \bibinfo {author} {\bibfnamefont {J.}~\bibnamefont {Skalski}},\ }\href
  {\doibase 10.1103/PhysRevC.83.054302} {\bibfield  {journal} {\bibinfo
  {journal} {Phys. Rev. C}\ }\textbf {\bibinfo {volume} {83}},\ \bibinfo
  {pages} {054302} (\bibinfo {year} {2011})}\BibitemShut {NoStop}%
\bibitem [{\citenamefont {Staszczak}\ \emph {et~al.}(2017)\citenamefont
  {Staszczak}, \citenamefont {Wong},\ and\ \citenamefont
  {Kosior}}]{Staszczak17}%
  \BibitemOpen
  \bibfield  {author} {\bibinfo {author} {\bibfnamefont {A.}~\bibnamefont
  {Staszczak}}, \bibinfo {author} {\bibfnamefont {C.-Y.}\ \bibnamefont {Wong}},
  \ and\ \bibinfo {author} {\bibfnamefont {A.}~\bibnamefont {Kosior}},\ }\href
  {\doibase 10.1103/PhysRevC.95.054315} {\bibfield  {journal} {\bibinfo
  {journal} {Phys. Rev. C}\ }\textbf {\bibinfo {volume} {95}},\ \bibinfo
  {pages} {054315} (\bibinfo {year} {2017})}\BibitemShut {NoStop}%
\bibitem [{\citenamefont {Borderie}\ \emph
  {et~al.}(1993{\natexlab{a}})\citenamefont {Borderie}, \citenamefont {Remaud},
  \citenamefont {Rivet},\ and\ \citenamefont {Sebille}}]{Borderie1993}%
  \BibitemOpen
  \bibfield  {author} {\bibinfo {author} {\bibfnamefont {B.}~\bibnamefont
  {Borderie}}, \bibinfo {author} {\bibfnamefont {B.}~\bibnamefont {Remaud}},
  \bibinfo {author} {\bibfnamefont {M.}~\bibnamefont {Rivet}}, \ and\ \bibinfo
  {author} {\bibfnamefont {F.}~\bibnamefont {Sebille}},\ }\href {\doibase
  http://dx.doi.org/10.1016/0370-2693(93)90628-U} {\bibfield  {journal}
  {\bibinfo  {journal} {Phys. Lett. B}\ }\textbf {\bibinfo {volume} {302}},\
  \bibinfo {pages} {15 } (\bibinfo {year} {1993}{\natexlab{a}})}\BibitemShut
  {NoStop}%
\bibitem [{\citenamefont {Borderie}\ \emph
  {et~al.}(1993{\natexlab{b}})\citenamefont {Borderie}, \citenamefont {Remaud},
  \citenamefont {Rivet},\ and\ \citenamefont {Sebille}}]{Borderie1993a}%
  \BibitemOpen
  \bibfield  {author} {\bibinfo {author} {\bibfnamefont {B.}~\bibnamefont
  {Borderie}}, \bibinfo {author} {\bibfnamefont {B.}~\bibnamefont {Remaud}},
  \bibinfo {author} {\bibfnamefont {M.}~\bibnamefont {Rivet}}, \ and\ \bibinfo
  {author} {\bibfnamefont {F.}~\bibnamefont {Sebille}},\ }\href {\doibase
  http://dx.doi.org/10.1016/0370-2693(93)90241-9} {\bibfield  {journal}
  {\bibinfo  {journal} {Phys. Lett. B}\ }\textbf {\bibinfo {volume} {307}},\
  \bibinfo {pages} {404} (\bibinfo {year} {1993}{\natexlab{b}})}\BibitemShut
  {NoStop}%
\bibitem [{\citenamefont {Bauer}\ \emph {et~al.}(1992)\citenamefont {Bauer},
  \citenamefont {Bertsch},\ and\ \citenamefont {Schulz}}]{Bauer1992}%
  \BibitemOpen
  \bibfield  {author} {\bibinfo {author} {\bibfnamefont {W.}~\bibnamefont
  {Bauer}}, \bibinfo {author} {\bibfnamefont {G.~F.}\ \bibnamefont {Bertsch}},
  \ and\ \bibinfo {author} {\bibfnamefont {H.}~\bibnamefont {Schulz}},\ }\href
  {\doibase 10.1103/PhysRevLett.69.1888} {\bibfield  {journal} {\bibinfo
  {journal} {Phys. Rev. Lett.}\ }\textbf {\bibinfo {volume} {69}},\ \bibinfo
  {pages} {1888} (\bibinfo {year} {1992})}\BibitemShut {NoStop}%
\bibitem [{\citenamefont {Xu}\ \emph {et~al.}(1994)\citenamefont {Xu},
  \citenamefont {Gagliardi}, \citenamefont {Tribble},\ and\ \citenamefont
  {Wong}}]{Xu1994}%
  \BibitemOpen
  \bibfield  {author} {\bibinfo {author} {\bibfnamefont {H.~M.}\ \bibnamefont
  {Xu}}, \bibinfo {author} {\bibfnamefont {C.~A.}\ \bibnamefont {Gagliardi}},
  \bibinfo {author} {\bibfnamefont {R.~E.}\ \bibnamefont {Tribble}}, \ and\
  \bibinfo {author} {\bibfnamefont {C.~Y.}\ \bibnamefont {Wong}},\ }\href
  {\doibase 10.1103/PhysRevC.49.R1778} {\bibfield  {journal} {\bibinfo
  {journal} {Phys. Rev. C}\ }\textbf {\bibinfo {volume} {49}},\ \bibinfo
  {pages} {R1778} (\bibinfo {year} {1994})}\BibitemShut {NoStop}%
\bibitem [{\citenamefont {Horowitz}\ and\ \citenamefont {Shen}(2008)}]{Hor08}%
  \BibitemOpen
  \bibfield  {author} {\bibinfo {author} {\bibfnamefont {C.~J.}\ \bibnamefont
  {Horowitz}}\ and\ \bibinfo {author} {\bibfnamefont {G.}~\bibnamefont
  {Shen}},\ }\href {\doibase 10.1103/PhysRevC.78.015801} {\bibfield  {journal}
  {\bibinfo  {journal} {Phys. Rev. C}\ }\textbf {\bibinfo {volume} {78}},\
  \bibinfo {pages} {015801} (\bibinfo {year} {2008})}\BibitemShut {NoStop}%
\bibitem [{\citenamefont {Wu}\ \emph {et~al.}(2014)\citenamefont {Wu},
  \citenamefont {Yao},\ and\ \citenamefont {Li}}]{Wu2014}%
  \BibitemOpen
  \bibfield  {author} {\bibinfo {author} {\bibfnamefont {X.~Y.}\ \bibnamefont
  {Wu}}, \bibinfo {author} {\bibfnamefont {J.~M.}\ \bibnamefont {Yao}}, \ and\
  \bibinfo {author} {\bibfnamefont {Z.~P.}\ \bibnamefont {Li}},\ }\href
  {\doibase 10.1103/PhysRevC.89.017304} {\bibfield  {journal} {\bibinfo
  {journal} {Phys. Rev. C}\ }\textbf {\bibinfo {volume} {89}},\ \bibinfo
  {pages} {017304} (\bibinfo {year} {2014})}\BibitemShut {NoStop}%
\bibitem [{\citenamefont {Shukla}\ and\ \citenamefont
  {\AA{}berg}(2014)}]{Shukla14}%
  \BibitemOpen
  \bibfield  {author} {\bibinfo {author} {\bibfnamefont {A.}~\bibnamefont
  {Shukla}}\ and\ \bibinfo {author} {\bibfnamefont {S.}~\bibnamefont
  {\AA{}berg}},\ }\href {\doibase 10.1103/PhysRevC.89.014329} {\bibfield
  {journal} {\bibinfo  {journal} {Phys. Rev. C}\ }\textbf {\bibinfo {volume}
  {89}},\ \bibinfo {pages} {014329} (\bibinfo {year} {2014})}\BibitemShut
  {NoStop}%
\bibitem [{\citenamefont {Friedrich}\ and\ \citenamefont
  {V{\"o}gler}(1982)}]{Fri82a}%
  \BibitemOpen
  \bibfield  {author} {\bibinfo {author} {\bibfnamefont {J.}~\bibnamefont
  {Friedrich}}\ and\ \bibinfo {author} {\bibfnamefont {N.}~\bibnamefont
  {V{\"o}gler}},\ }\href@noop {} {\bibfield  {journal} {\bibinfo  {journal}
  {Nucl. Phys. A}\ }\textbf {\bibinfo {volume} {373}},\ \bibinfo {pages} {192}
  (\bibinfo {year} {1982})}\BibitemShut {NoStop}%
\bibitem [{\citenamefont {Bender}\ \emph {et~al.}(2003)\citenamefont {Bender},
  \citenamefont {Heenen},\ and\ \citenamefont {Reinhard}}]{bender2003self}%
  \BibitemOpen
  \bibfield  {author} {\bibinfo {author} {\bibfnamefont {M.}~\bibnamefont
  {Bender}}, \bibinfo {author} {\bibfnamefont {P.-H.}\ \bibnamefont {Heenen}},
  \ and\ \bibinfo {author} {\bibfnamefont {P.-G.}\ \bibnamefont {Reinhard}},\
  }\href@noop {} {\bibfield  {journal} {\bibinfo  {journal} {Rev. Mod. Phys.}\
  }\textbf {\bibinfo {volume} {75}},\ \bibinfo {pages} {121} (\bibinfo {year}
  {2003})}\BibitemShut {NoStop}%
\bibitem [{\citenamefont {Kl{\"{u}}pfel}\ \emph {et~al.}(2009)\citenamefont
  {Kl{\"{u}}pfel}, \citenamefont {Reinhard}, \citenamefont {B{\"{u}}rvenich},\
  and\ \citenamefont {Maruhn}}]{Kluepfel2009}%
  \BibitemOpen
  \bibfield  {author} {\bibinfo {author} {\bibfnamefont {P.}~\bibnamefont
  {Kl{\"{u}}pfel}}, \bibinfo {author} {\bibfnamefont {P.~G.}\ \bibnamefont
  {Reinhard}}, \bibinfo {author} {\bibfnamefont {T.~J.}\ \bibnamefont
  {B{\"{u}}rvenich}}, \ and\ \bibinfo {author} {\bibfnamefont {J.~A.}\
  \bibnamefont {Maruhn}},\ }\href@noop {} {\bibfield  {journal} {\bibinfo
  {journal} {Phys. Rev. C}\ }\textbf {\bibinfo {volume} {79}},\ \bibinfo
  {pages} {034310} (\bibinfo {year} {2009})}\BibitemShut {NoStop}%
\bibitem [{\citenamefont {Chabanat}\ \emph {et~al.}(1998)\citenamefont
  {Chabanat}, \citenamefont {Bonche}, \citenamefont {Haensel}, \citenamefont
  {Meyer},\ and\ \citenamefont {Schaeffer}}]{Chabanat}%
  \BibitemOpen
  \bibfield  {author} {\bibinfo {author} {\bibfnamefont {E.}~\bibnamefont
  {Chabanat}}, \bibinfo {author} {\bibfnamefont {P.}~\bibnamefont {Bonche}},
  \bibinfo {author} {\bibfnamefont {P.}~\bibnamefont {Haensel}}, \bibinfo
  {author} {\bibfnamefont {J.}~\bibnamefont {Meyer}}, \ and\ \bibinfo {author}
  {\bibfnamefont {R.}~\bibnamefont {Schaeffer}},\ }\href@noop {} {\bibfield
  {journal} {\bibinfo  {journal} {Nucl. Phys. A}\ }\textbf {\bibinfo {volume}
  {635}},\ \bibinfo {pages} {231 } (\bibinfo {year} {1998})}\BibitemShut
  {NoStop}%
\bibitem [{\citenamefont {Kortelainen}\ \emph {et~al.}(2012)\citenamefont
  {Kortelainen}, \citenamefont {McDonnell}, \citenamefont {Nazarewicz},
  \citenamefont {Reinhard}, \citenamefont {Sarich}, \citenamefont {Schunck},
  \citenamefont {Stoitsov},\ and\ \citenamefont {Wild}}]{Kortelainen2012b}%
  \BibitemOpen
  \bibfield  {author} {\bibinfo {author} {\bibfnamefont {M.}~\bibnamefont
  {Kortelainen}}, \bibinfo {author} {\bibfnamefont {J.}~\bibnamefont
  {McDonnell}}, \bibinfo {author} {\bibfnamefont {W.}~\bibnamefont
  {Nazarewicz}}, \bibinfo {author} {\bibfnamefont {P.-G.}\ \bibnamefont
  {Reinhard}}, \bibinfo {author} {\bibfnamefont {J.}~\bibnamefont {Sarich}},
  \bibinfo {author} {\bibfnamefont {N.}~\bibnamefont {Schunck}}, \bibinfo
  {author} {\bibfnamefont {M.~V.}\ \bibnamefont {Stoitsov}}, \ and\ \bibinfo
  {author} {\bibfnamefont {S.~M.}\ \bibnamefont {Wild}},\ }\href {\doibase
  10.1103/PhysRevC.85.024304} {\bibfield  {journal} {\bibinfo  {journal} {Phys.
  Rev. C}\ }\textbf {\bibinfo {volume} {85}},\ \bibinfo {pages} {024304}
  (\bibinfo {year} {2012})}\BibitemShut {NoStop}%
\bibitem [{\citenamefont {Bonche}\ \emph {et~al.}(1985)\citenamefont {Bonche},
  \citenamefont {Flocard}, \citenamefont {Heenen}, \citenamefont {Krieger},\
  and\ \citenamefont {Weiss}}]{Bon85a}%
  \BibitemOpen
  \bibfield  {author} {\bibinfo {author} {\bibfnamefont {P.}~\bibnamefont
  {Bonche}}, \bibinfo {author} {\bibfnamefont {H.}~\bibnamefont {Flocard}},
  \bibinfo {author} {\bibfnamefont {P.-H.}\ \bibnamefont {Heenen}}, \bibinfo
  {author} {\bibfnamefont {S.~J.}\ \bibnamefont {Krieger}}, \ and\ \bibinfo
  {author} {\bibfnamefont {M.~S.}\ \bibnamefont {Weiss}},\ }\href@noop {}
  {\bibfield  {journal} {\bibinfo  {journal} {Nucl. Phys. A}\ }\textbf
  {\bibinfo {volume} {443}},\ \bibinfo {pages} {39} (\bibinfo {year}
  {1985})}\BibitemShut {NoStop}%
\bibitem [{\citenamefont {Krieger}\ \emph {et~al.}(1990)\citenamefont
  {Krieger}, \citenamefont {Bonche}, \citenamefont {Flocard}, \citenamefont
  {Quentin},\ and\ \citenamefont {Weiss}}]{Kri90a}%
  \BibitemOpen
  \bibfield  {author} {\bibinfo {author} {\bibfnamefont {S.~J.}\ \bibnamefont
  {Krieger}}, \bibinfo {author} {\bibfnamefont {P.}~\bibnamefont {Bonche}},
  \bibinfo {author} {\bibfnamefont {H.}~\bibnamefont {Flocard}}, \bibinfo
  {author} {\bibfnamefont {P.}~\bibnamefont {Quentin}}, \ and\ \bibinfo
  {author} {\bibfnamefont {M.~S.}\ \bibnamefont {Weiss}},\ }\href@noop {}
  {\bibfield  {journal} {\bibinfo  {journal} {Nucl. Phys. A}\ }\textbf
  {\bibinfo {volume} {517}},\ \bibinfo {pages} {275} (\bibinfo {year}
  {1990})}\BibitemShut {NoStop}%
\bibitem [{\citenamefont {Bender}\ \emph {et~al.}(2000)\citenamefont {Bender},
  \citenamefont {Rutz}, \citenamefont {Reinhard},\ and\ \citenamefont
  {Maruhn}}]{Ben00}%
  \BibitemOpen
  \bibfield  {author} {\bibinfo {author} {\bibfnamefont {M.}~\bibnamefont
  {Bender}}, \bibinfo {author} {\bibfnamefont {K.}~\bibnamefont {Rutz}},
  \bibinfo {author} {\bibfnamefont {P.-G.}\ \bibnamefont {Reinhard}}, \ and\
  \bibinfo {author} {\bibfnamefont {J.~A.}\ \bibnamefont {Maruhn}},\
  }\href@noop {} {\bibfield  {journal} {\bibinfo  {journal} {Eur. Phys. J. A}\
  }\textbf {\bibinfo {volume} {8}},\ \bibinfo {pages} {59} (\bibinfo {year}
  {2000})}\BibitemShut {NoStop}%
\bibitem [{\citenamefont {Dobaczewski}\ \emph {et~al.}(2014)\citenamefont
  {Dobaczewski}, \citenamefont {Nazarewicz},\ and\ \citenamefont
  {Reinhard}}]{Dobaczewski2014}%
  \BibitemOpen
  \bibfield  {author} {\bibinfo {author} {\bibfnamefont {J.}~\bibnamefont
  {Dobaczewski}}, \bibinfo {author} {\bibfnamefont {W.}~\bibnamefont
  {Nazarewicz}}, \ and\ \bibinfo {author} {\bibfnamefont {P.-G.}\ \bibnamefont
  {Reinhard}},\ }\href {\doibase 10.1088/0954-3899/41/7/074001} {\bibfield
  {journal} {\bibinfo  {journal} {J. Phys. G}\ }\textbf {\bibinfo {volume}
  {41}},\ \bibinfo {pages} {074001} (\bibinfo {year} {2014})}\BibitemShut
  {NoStop}%
\bibitem [{\citenamefont {Glantz}\ \emph {et~al.}(1990)\citenamefont {Glantz},
  \citenamefont {Slinker},\ and\ \citenamefont {Neilands}}]{Glantz}%
  \BibitemOpen
  \bibfield  {author} {\bibinfo {author} {\bibfnamefont {S.~A.}\ \bibnamefont
  {Glantz}}, \bibinfo {author} {\bibfnamefont {B.~K.}\ \bibnamefont {Slinker}},
  \ and\ \bibinfo {author} {\bibfnamefont {T.~B.}\ \bibnamefont {Neilands}},\
  }\href@noop {} {\emph {\bibinfo {title} {Primer of Applied Regression \&
  Analysis of Variance}}}\ (\bibinfo  {publisher} {McGraw Hill},\ \bibinfo
  {year} {1990})\BibitemShut {NoStop}%
\bibitem [{\citenamefont {Reinhard}(2016)}]{Reinhard16}%
  \BibitemOpen
  \bibfield  {author} {\bibinfo {author} {\bibfnamefont {P.-G.}\ \bibnamefont
  {Reinhard}},\ }\href {http://stacks.iop.org/1402-4896/91/i=2/a=023002}
  {\bibfield  {journal} {\bibinfo  {journal} {Phys. Scr.}\ }\textbf {\bibinfo
  {volume} {91}},\ \bibinfo {pages} {023002} (\bibinfo {year}
  {2016})}\BibitemShut {NoStop}%
\bibitem [{\citenamefont {Allison}(1998)}]{Allison}%
  \BibitemOpen
  \bibfield  {author} {\bibinfo {author} {\bibfnamefont {P.~D.}\ \bibnamefont
  {Allison}},\ }\href@noop {} {\emph {\bibinfo {title} {Multiple Regression: A
  Primer}}}\ (\bibinfo  {publisher} {Sage Publications},\ \bibinfo {year}
  {1998})\BibitemShut {NoStop}%
\bibitem [{\citenamefont {Oganessian}\ \emph {et~al.}(2006)\citenamefont
  {Oganessian} \emph {et~al.}}]{Oga1}%
  \BibitemOpen
  \bibfield  {author} {\bibinfo {author} {\bibfnamefont {Y.~T.}\ \bibnamefont
  {Oganessian}} \emph {et~al.},\ }\href {\doibase 10.1103/PhysRevC.74.044602}
  {\bibfield  {journal} {\bibinfo  {journal} {Phys. Rev. C}\ }\textbf {\bibinfo
  {volume} {74}},\ \bibinfo {pages} {044602} (\bibinfo {year}
  {2006})}\BibitemShut {NoStop}%
\bibitem [{\citenamefont {\'Cwiok}\ \emph {et~al.}(2005)\citenamefont
  {\'Cwiok}, \citenamefont {Heenen},\ and\ \citenamefont
  {Nazarewicz}}]{SHENature}%
  \BibitemOpen
  \bibfield  {author} {\bibinfo {author} {\bibfnamefont {S.}~\bibnamefont
  {\'Cwiok}}, \bibinfo {author} {\bibfnamefont {P.~H.}\ \bibnamefont {Heenen}},
  \ and\ \bibinfo {author} {\bibfnamefont {W.}~\bibnamefont {Nazarewicz}},\
  }\href {http://dx.doi.org/10.1038/nature03336} {\bibfield  {journal}
  {\bibinfo  {journal} {Nature}\ }\textbf {\bibinfo {volume} {433}},\ \bibinfo
  {pages} {705} (\bibinfo {year} {2005})}\BibitemShut {NoStop}%
\bibitem [{\citenamefont {Heenen}\ \emph {et~al.}(2015)\citenamefont {Heenen},
  \citenamefont {Skalski}, \citenamefont {Staszczak},\ and\ \citenamefont
  {Vretenar}}]{Heenen15}%
  \BibitemOpen
  \bibfield  {author} {\bibinfo {author} {\bibfnamefont {P.-H.}\ \bibnamefont
  {Heenen}}, \bibinfo {author} {\bibfnamefont {J.}~\bibnamefont {Skalski}},
  \bibinfo {author} {\bibfnamefont {A.}~\bibnamefont {Staszczak}}, \ and\
  \bibinfo {author} {\bibfnamefont {D.}~\bibnamefont {Vretenar}},\ }\href
  {\doibase http://dx.doi.org/10.1016/j.nuclphysa.2015.07.016} {\bibfield
  {journal} {\bibinfo  {journal} {Nucl. Phys. A}\ }\textbf {\bibinfo {volume}
  {944}},\ \bibinfo {pages} {415 } (\bibinfo {year} {2015})}\BibitemShut
  {NoStop}%
\bibitem [{\citenamefont {Kortelainen}\ \emph {et~al.}(2010)\citenamefont
  {Kortelainen}, \citenamefont {Lesinski}, \citenamefont {Mor\'e},
  \citenamefont {Nazarewicz}, \citenamefont {Sarich}, \citenamefont {Schunck},
  \citenamefont {Stoitsov},\ and\ \citenamefont {Wild}}]{Kortelainen2010}%
  \BibitemOpen
  \bibfield  {author} {\bibinfo {author} {\bibfnamefont {M.}~\bibnamefont
  {Kortelainen}}, \bibinfo {author} {\bibfnamefont {T.}~\bibnamefont
  {Lesinski}}, \bibinfo {author} {\bibfnamefont {J.}~\bibnamefont {Mor\'e}},
  \bibinfo {author} {\bibfnamefont {W.}~\bibnamefont {Nazarewicz}}, \bibinfo
  {author} {\bibfnamefont {J.}~\bibnamefont {Sarich}}, \bibinfo {author}
  {\bibfnamefont {N.}~\bibnamefont {Schunck}}, \bibinfo {author} {\bibfnamefont
  {M.~V.}\ \bibnamefont {Stoitsov}}, \ and\ \bibinfo {author} {\bibfnamefont
  {S.}~\bibnamefont {Wild}},\ }\href {\doibase 10.1103/PhysRevC.82.024313}
  {\bibfield  {journal} {\bibinfo  {journal} {Phys. Rev. C}\ }\textbf {\bibinfo
  {volume} {82}},\ \bibinfo {pages} {024313} (\bibinfo {year}
  {2010})}\BibitemShut {NoStop}%
\bibitem [{\citenamefont {Reinhard}\ \emph {et~al.}(2006)\citenamefont
  {Reinhard}, \citenamefont {Bender}, \citenamefont {Nazarewicz},\ and\
  \citenamefont {Vertse}}]{Leptodermous}%
  \BibitemOpen
  \bibfield  {author} {\bibinfo {author} {\bibfnamefont {P.-G.}\ \bibnamefont
  {Reinhard}}, \bibinfo {author} {\bibfnamefont {M.}~\bibnamefont {Bender}},
  \bibinfo {author} {\bibfnamefont {W.}~\bibnamefont {Nazarewicz}}, \ and\
  \bibinfo {author} {\bibfnamefont {T.}~\bibnamefont {Vertse}},\ }\href
  {\doibase 10.1103/PhysRevC.73.014309} {\bibfield  {journal} {\bibinfo
  {journal} {Phys. Rev. C}\ }\textbf {\bibinfo {volume} {73}},\ \bibinfo
  {pages} {014309} (\bibinfo {year} {2006})}\BibitemShut {NoStop}%
\end{thebibliography}%

\end{document}